\documentclass[twocolumn]{aastex631}
\usepackage{amsmath,amsthm,amssymb}

\begin{document}

%\title{How Resistivity Affects the Periodicity of Oscillatory Reconnection}

\title{The Effect of Resistivity on the Periodicity of Oscillatory Reconnection}

\author[0009-0008-7874-2449]{Jordan Talbot}
\affiliation{Department of Mathematics, Physics \& Electrical Engineering, Northumbria University, Newcastle upon Tyne NE1 8ST, UK}

\author[0000-0002-7863-624X]{James A. McLaughlin}
\affiliation{Department of Mathematics, Physics \& Electrical Engineering, Northumbria University, Newcastle upon Tyne NE1 8ST, UK}

\author[0000-0002-5915-697X]{Gert J.J. Botha}
\affiliation{Department of Mathematics, Physics \& Electrical Engineering, Northumbria University, Newcastle upon Tyne NE1 8ST, UK}

\author[0009-0009-5929-9776]{Mark Hancock}
\affiliation{Department of Mathematics, Physics \& Electrical Engineering, Northumbria University, Newcastle upon Tyne NE1 8ST, UK}

%%%%%%%%%%%%%%%%%%%%%%%%%%%%%%%%%%%%%%%%%%%%%

%\begin{abstract}
%\textit{Context:} MHD waves and magnetic null points are both ubiquitous within the solar atmosphere and one mechanism for their interaction is oscillatory reconnection.\\
%\textit{Aim:} In this study, the oscillatory reconnection system is investigated for different levels of resistivity, focusing on the period of the system and the currents produced. \\
%\textit{Method:} Resistive MHD simulations are ran, with the use of the Lare2d code, for different levels of resistivity. Using three methods; wavelet analysis, Fourier transform and ANOVA, the affect of resitivity on period of the system is investigated. The affect of resistivity on the currents in the system is also investigated, looking at the amplitude and decay rates of current oscillations and the levels of ohmic heating. \\
%\textit{Results:} It is found that there is an independence between the level of background resistivity and the period of the oscillatory reconnection system. It is also found that resistivity has a direct affect on the currents in the system, including the ohmic heating.\\
%\textit{Conclusions:} Periodicity of oscillatory reconnection is independent of background resistivity levels. However, current density amplitude, current density oscillation decay rate and levels of ohmic heating are dependent on resistivity.\\
%\end{abstract}

%%%%%%%%%%%%%%%%%%%%%%%%%%%%%%%%%%%%%%%%%

\begin{abstract}
The oscillatory reconnection mechanism is investigated for a parameter study of eight orders of magnitude of resistivity, with a particular interest in the evolution of the oscillating current density at the null point and its associated periodicity. The resistive, nonlinear MHD simulations are solved in 2.5D for different levels of resistivity. Three methods (wavelet analysis, Fourier transform and ANOVA) are used to investigate the effect of resistivity versus resultant period. It is found that there is an independence between the level of background resistivity and the period of the oscillatory reconnection mechanism. Conversely, it is found that resistivity has a significant effect on the maximum amplitude of the current density and the nature of its decay rate, as well as the magnitude of ohmic heating at the null. 
\end{abstract}


%%%%%%%%%%%%%%%%%%%%%%%%%%%%%%%%%%%%%%%

\section{Introduction} \label{sec:intro}

The solar magnetic field is a highly dynamic system, with motions constantly creating waves and flows within the field. Furthermore, a large number of regions with different magnetic polarities can lead to the formation of null points. Magnetic null points are singularities within the magnetic field where the field strength, and thus the Alfv\'en speed, are zero. Studies have shown, through the use of magnetic field extrapolations from magnetogram observations, that magnetic null points are ubiquitous within the solar atmosphere \citep[e.g.][]{2005LRSP....2....7L, 2008A&A...484L..47R}. Due to this ubiquity, a large number of studies have been made on the role of null points, in particular how they interact with the motion of magnetohydrodynamic (MHD) waves through wave propagation  \citep[e.g.][]{2004A&A...420.1129M, 2005A&A...435..313M, 2006A&A...452..603M,2011SSRv..158..205M,2022ApJ...924..126S, 2023ApJ...944...72S} and mode conversion \citep[e.g.][]{2006A&A...459..641M,2012A&A...545A...9T,2019ApJ...879..127T,2021SoPh..296...97P,2022A&A...660A..21Y} but also as sources of slow, fast and Alfv\'en waves  \citep[e.g.][]{2014SoPh..289.3043L,2017ApJ...834...62K,2017ApJ...844....2T}{}{}.

%Null points can also be the location of magnetic reconnection events.

Magnetic reconnection is a fundamental process in plasma physics where the topology of the magnetic field is rearranged. Strong currents are produced during reconnection events, which, due to electrical resistivity, allow magnetic fields lines to diffuse and change their connectivity \citep[e.g.][]{1957JGR....62..509P, 1958IAUS....6..123S, 1964NASSP..50..425P}. In 2D, reconnection is understood to occur mainly at null points where it can be driven by a number of processes such as variations in the current density \cite[]{2000mrmt.conf.....P}. Reconnection events can lead to heating of the surrounding plasma through ohmic heating and shock heating, as well as causing particle acceleration and localized motion. 3D reconnection is a fundamentally different system and readers are directed to \cite{2003JGRA..108.1285P}, \cite{2011AdSpR..47.1508P} and \cite{2022LRSP...19....1P} for comprehensive reviews.

Generally, reconnection is studied using steady state models, either the Sweet-Parker model \citep[]{1957JGR....62..509P, 1958IAUS....6..123S} or the Petschek model \cite[]{1964NASSP..50..425P}. However, magnetic reconnection is a leading theory behind the origin of dynamic solar events such as solar flares \citep[e.g.][]{1999Ap&SS.264..129S, 2013NatPh...9..489S}{}{} and thus a time-dependent form of reconnection is more applicable to such dynamic events. One such time-dependent form is that of oscillatory reconnection (OR).

%%%%%%%%%%%%%%%%%%%%%%%%%%%%%%%%%%%%%%%%%%%%%%%%%%%%%%%%
%\begin{figure}[t]
%    \centering
%    \includegraphics[width=0.5\textwidth]{Mag_Field.eps}
%        \includegraphics[width=0.4\textwidth]{Mag_Field.eps}
%    \caption{Equilibrium magnetic field (black lines) defined by Equation (\ref{MAGEQ}) where the orange lines are the separatrices which split the domain into regions of different connectivity and the arrows indicate the direction of the magnetic field. The magnetic null point is located at $(x=0,y=0)$, where the separatrices intersect.}
%    \label{MAG_FIELD}
%\end{figure}
%%%%%%%%%%%%%%%%%%%%%%%%%%%%%%%%%%%%%%%%%%%%%%%%%%%%%%%%

%%%%%%%%%%%%%%%%%%%%%%%%%%%%%%%%%%%%%%%%%%%%%%%%%%%%%
\begin{figure*}[t]
    \centering
    \includegraphics[width=2in]{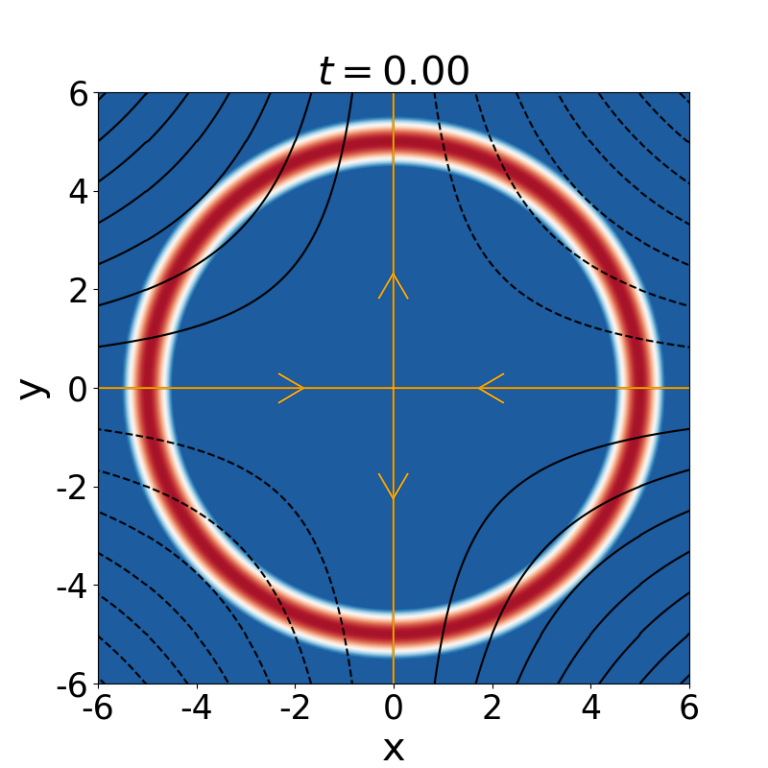} 
    \includegraphics[width=2in]{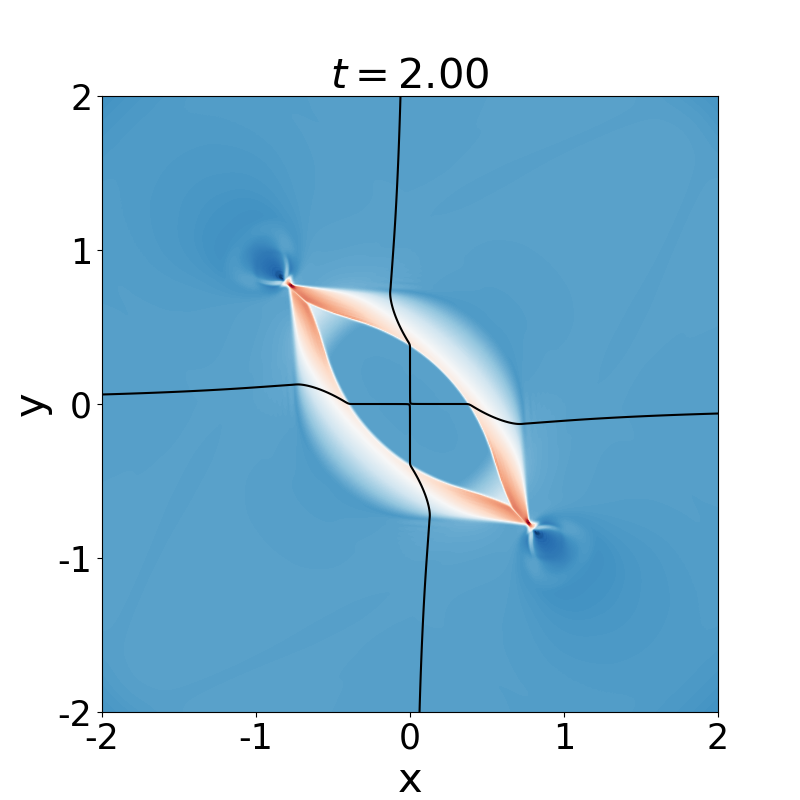} 
    \includegraphics[width=2in]{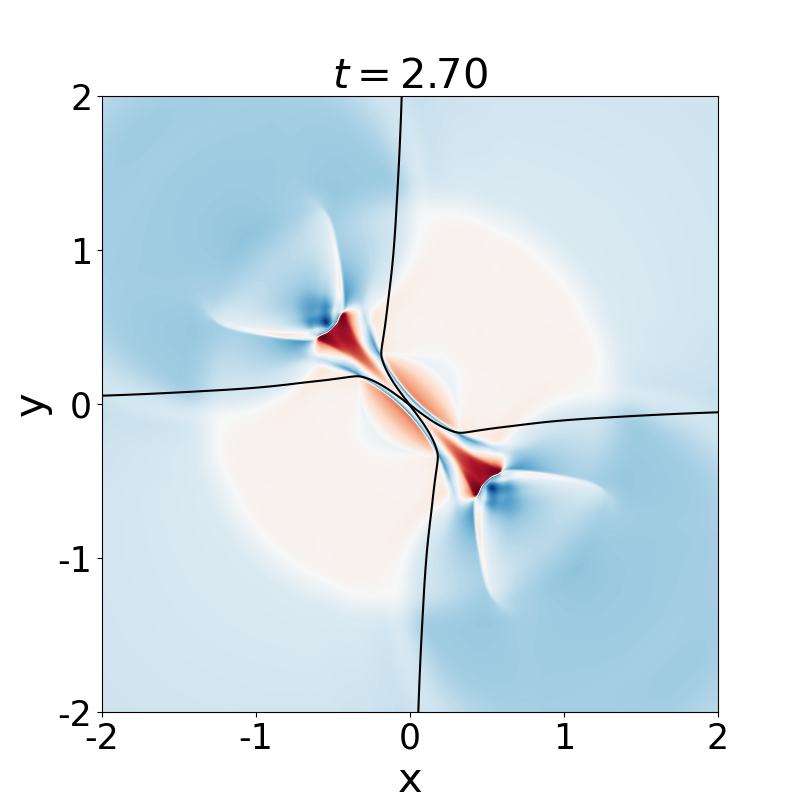}  \\
    \includegraphics[width=2in]{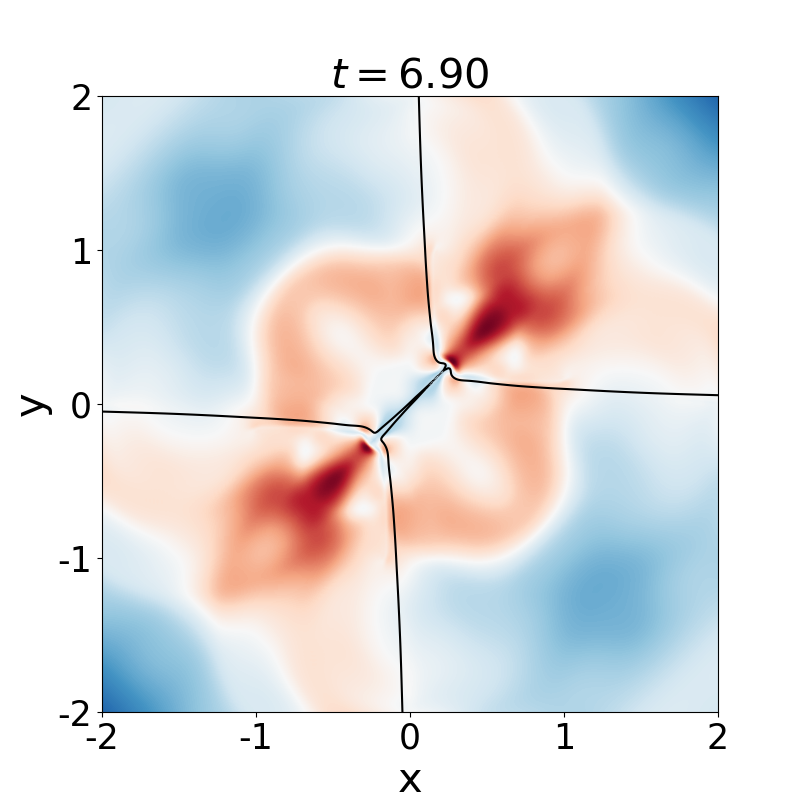} 
    \includegraphics[width=2in]{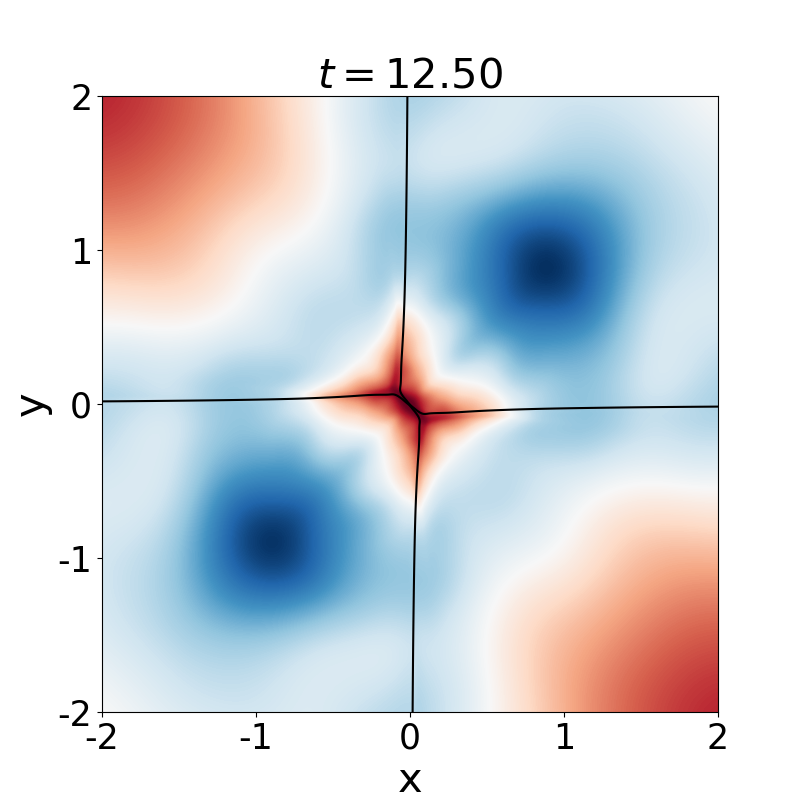} 
    \includegraphics[width=2in]{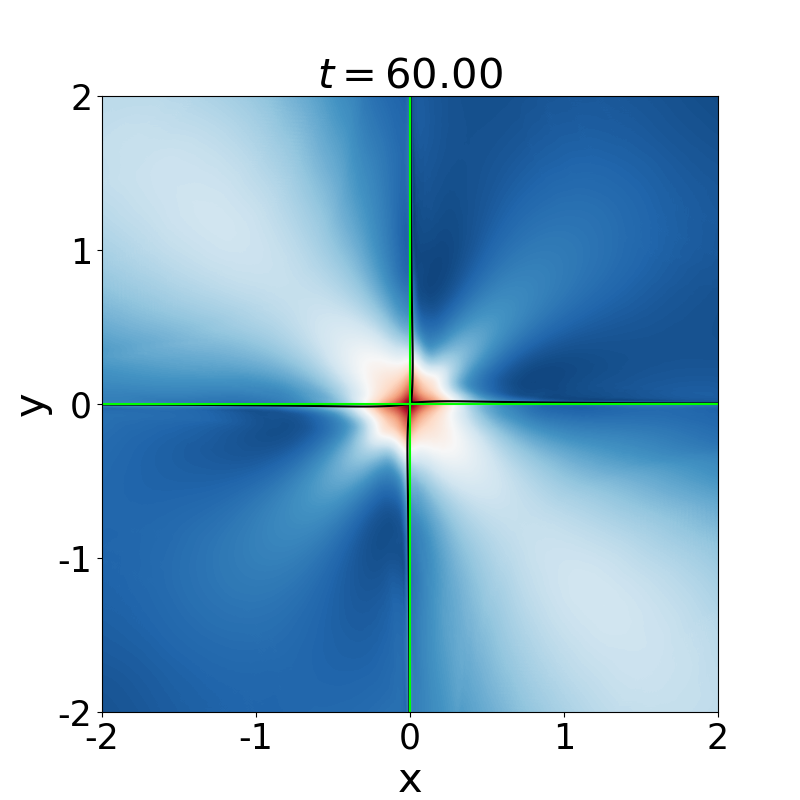}
    \caption{Contours of $v_{\perp}$ at times $t=0$, $2.0$, $2.7$,  $6.9$, $12.5$ and $ 60$. $t=0$ shows the initial pulse of $v_{\perp}$ (blue-to-red contour) and the equilibrium magnetic field, defined by Equation (\ref{MAGEQ}), where the black lines denote the magnetic field lines and the orange lines denote the seperatrices. The arrows indicate the direction of the magnetic field. $t=2.0$ shows the formation of shocks on the wave pulse, which bend the magnetic field lines away from the normal. $t=2.7$ shows the formation of the first current sheet. $t=6.9$ and $t=12.5$ show the change in the orientation of the current sheet. The black lines denote the  evolving separatrices. $t=60.0$ is the final snapshot from the simulation where  the green line denotes the  original separatrices at $t=0$.}
    \label{SIMSNAPS}
\end{figure*}
%%%%%%%%%%%%%%%%%%%%%%%%%%%%%%%%%%%%%%%%%%%%%%%%%%%%%

OR was first studied by \cite{1991ApJ...371L..41C} who found that a null point, when perturbed from equilibrium and subsequently allowed to relax, could produce a reconnecting current sheet which oscillated about its original equilibrium orientation. It was then found by \cite{2009A&A...493..227M} that OR could be driven by a nonlinear fast magnetoacoustic wave which shocks and then collapses the null. OR has been studied in 3D by \cite{2017ApJ...844....2T} and more recently it has been found that it can be driven through the merging of magnetic flux ropes \cite[]{2022MNRAS.513.5224S}. \cite{2022ApJ...925..195K} simulated OR within 1\:MK coronal plasma.

In particular, OR has become one of the leading contenders for the origin of quasi-periodic pulsations from solar flares \citep[][]{2018SSRv..214...45M, 2021SSRv..217...66Z,2022NatCo..13.7680K}. OR has also become a potential theoretical explanation for the origin of other periodic solar phenomena, including periodicities within solar jets \cite[]{2019ApJ...874..146H}, quasi-periodic flows  \cite[]{2012ApJ...749...30M},  periodicities within Type III radio bursts \cite[]{2021A&A...650A...6C}, filament formation \cite[]{2023ApJ...944..161S}, flux rope formation and disappearance \cite[]{2019ApJ...874L..27X}, stellar flare oscillations \cite[]{2018MNRAS.475.2842D}, and a possible explanation for spikes and switchbacks observed by Parker Solar Probe \cite[]{2021ApJ...913L..14H}.

%and has also been muted as  and 

In recent years, a number of studies have investigated the OR mechanism further to understand how the process is affected by various changes in parameters. In particular, \cite{2022ApJ...925..195K,2022ApJ...933..142K} carried out a comprehensive study into how the period of the OR mechanism was affected by a number of environmental variables, such as temperature, density, background magnetic field strength, as well as the properties of the initial wave pulse. These studies led to the derivation of an empirical formula for the period of the OR oscillations in the electric current density and also led to the possibility of using OR as a plasma diagnostic in the corona \cite[]{2023ApJ...943..131K}.

In this paper, we will use a similar setup to that of \cite{2009A&A...493..227M} and study how variations in the levels of background resistivity affect the OR mechanism. We shall focus on how a change in resistivity, across multiple orders of magnitude, affects the current density produced at the location of the magnetic null point. In Section \ref{sec:setup} we describe our numerical approach to this study, outlining our physical domain, numerical setup and the code used. The results of this parameter study are presented in Section \ref{sec:results}, with a focus on the effect on the amplitude of the oscillations of the current density (Section \ref{sec:jzamp}), the effect on the period in the system (Section \ref{period}), the effect on the decay rate of the current density oscillations (Section \ref{sec:decay}) and the corresponding ohmic heating produced (Section \ref{sec:ohm}). In Section \ref{sec:conclusions} we present our conclusions and general discussion resulting from this study. Appendix \ref{Fourier} reports on an additional study on the period of the system using Fourier Spectra, with Appendix \ref{SUPPLOTS} showing the results of simulations complementing those in Section \ref{sec:results}.

%%%%%%%%%%%%%%%%%%%%%%%%%%%%%%%%%%%%%%%%%%%%
\begin{figure*}[t]
    \centering
    \includegraphics[width=\textwidth]{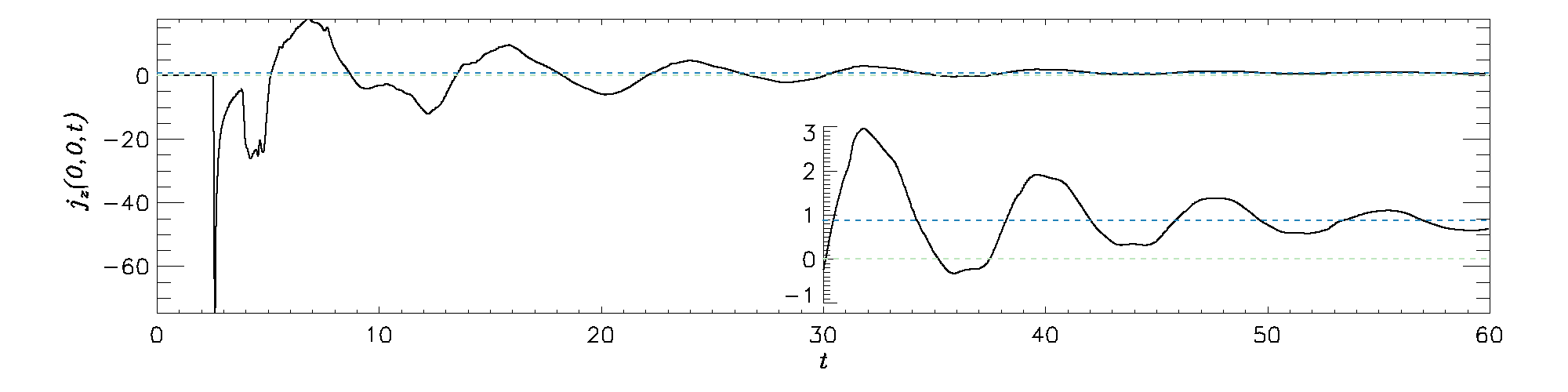}
    \caption{Plot of $j_z(0,0,t)$ at the null point against time for baseline simulation ($\eta=10^{-4}$). The green dashed line denotes $j_z = 0$ with the blue dashed line denoting $j_z = 0.87$, which is the value that the $j_z$ profile tends to. }
    \label{JZ}
\end{figure*}
%%%%%%%%%%%%%%%%%%%%%%%%%%%%%%%%%%%%%%%%%%%%

\section{Numerical Setup}\label{sec:setup}

\subsection{Governing Equations}
In this study the Lare2d code \cite[]{2001JCoPh.171..151A} is used to solve the resistive magnetohydrodynamic (MHD) equations in 2.5D, which are defined as:

\begin{eqnarray}
    \frac{\partial\rho}{\partial t} &=& -\nabla\cdot(\rho \mathbf{v})\, , \\
    \frac{D\mathbf{v}}{Dt} &=& \frac{1}{\rho}\mathbf{j}\times\mathbf{B} - \frac{1}{\rho}\nabla P \label{MOME} \, ,\\
    \frac{\partial\mathbf{B}}{\partial t} &=& -\nabla\times\mathbf{E}\, , \\
    \frac{D\epsilon}{Dt} &=& -\frac{P}{\rho}\nabla\cdot\mathbf{v} + \frac{\eta}{\rho}j^2 \, ,\\
    \mathbf{E} + \mathbf{v}\times\mathbf{B} &=& \eta\mathbf{j} \label{OHM} \, ,\\
    \nabla\times\mathbf{B} &=& \mu_0\mathbf{j}\, , \label{VXB}
\end{eqnarray}
where $\rho$ is density, $P$ is plasma pressure, $\mathbf{v}$ is velocity, $\mathbf{B}$ is the magnetic field, $\mathbf{E}$ is the electric field, $\mathbf{j}$ is the electric current density, $\mu_0$ is the magnetic permeability, $\epsilon$ is the specific internal energy density and $\eta$ is resistivity. In these equations $\frac{D}{Dt}$ is the advective derivative.

These equations can then be non-dimensionalised by letting: $\mathbf{v}=v_o\mathbf{v}^*$, $\mathbf{B} = B_0\mathbf{B}^*$, $x=Lx^*$, $y=Ly^*$, $\mathbf{j}=j_0\mathbf{j}^*$, $\rho=\rho_0\rho^*$, $P=P_0P^*$, $\mathbf{E} = E_0\mathbf{E}^*$, $\eta = \eta_0\eta^*$ and $t=t_0t^*$. Here $^*$ denotes the dimensionless parameter and $v_0$, $B_0$, $L$, $j_0$, $\rho_0$, $P_0$, $E_0$, $\eta_0$ and $t_0$ correspond to constants with the dimensions of the variable they are non-dimensionalising. Due to the scale free nature of our equilibrium magnetic field (Equation \ref{MAGEQ}), there is freedom in the choice of these constant values. For the remainder of this paper, the $^*$ is dropped from notation, with all variables being non-dimensionalised. Variables, such as velocity and time, can be re-dimensionalised through multiplication with their respective dimensional constants, such as $v_0 = B_0/\sqrt{\mu_0\rho_0}$ and $t_0 = L/v_0$.

The Lare2d code utilises artificial shock viscosities ($\nu_1$ and $\nu_2$), which introduce a level of dissipation at steep gradients, allowing the accurate capture of MHD shocks \cite[see][for details]{1998JCoPh.144...70C}. The values that are chosen in this study for these viscosities are $\nu_1 = 0.5$ and $\nu_2=0.1$.

%%%%%%%%%%%%%%%%%%%%%%%%%%%%%%%%%%%

\subsection{Initial Conditions}

The computational domain is defined as $[x,y] \in \left[-97.5,97.5\right]$ with a numerical resolution of $2560\times2560$ grid cells. A stretched grid is implemented within this domain in both the $x$ and $y$ directions, allowing for a uniform numerical resolution of $1280\times1280$ within the domain $[x,y] \in \left[-5,5\right]$. The implementation of this stretched grid allows for the uniform grid to be around the area of interest, the null point, but allows for the boundaries to be a significant distance away.

The system is initially defined as a cold plasma with a uniform density where the equilibrium magnetic field is defined similar to that of \cite{2009A&A...493..227M} but with a $\pi/4$ rotation:
\begin{equation}\label{MAGEQ}
    \mathbf{B}_0 = \frac{B}{L}\left(x,-y,0\right)\, ,
\end{equation}
which is defined so that the solenoidal condition $\nabla\cdot\mathbf{B}=0$ holds. It is worth noting here that this magnetic field becomes unphysical far from the null point due to the magnetic field strength tending towards unphysically large values. It has been shown in \cite{2006A&A...452..603M} that for equilibrium magnetic fields where this unphysical nature at large distances is removed, the key underlying physics near the null still hold.

From this magnetic field, the vector potential $\mathbf{A}$ can be calculated, where $\mathbf{B} = \nabla\times\mathbf{A}$. In the 2D coordinate system chosen, this equation becomes $\mathbf{A}=A_z\mathbf{\hat{z}}$, with the equilibrium vector potential defined as:
\begin{equation}
    \mathbf{A}_0 = A_0\mathbf{\hat{z}} = -xy\mathbf{\hat{z}}\;.
\end{equation}
This initial magnetic field can be seen in Figure \ref{SIMSNAPS} for $t=0.0$ along with its subsequent evolution.

In this system two main parameters related to the velocity are of interest, namely the parallel, $v_{\parallel} = \mathbf{v}\cdot\mathbf{B}$, and the perpendicular, $v_{\perp} = \mathbf{v}\times\mathbf{B} \cdot \mathbf{\hat{z}}$, components.

The initial wave pulse is defined in the perpendicular component, $v_{\perp}$, as a circular pulse around the null, as in \cite{2009A&A...493..227M} and \cite{2022ApJ...933..142K}. This initial condition, coupled with setting the parallel component $v_{\parallel}$ to zero, results in the initial wave pulse being a purely fast magnetoacoustic wave, defined by: 
\begin{eqnarray}
    v_{\perp}\left(x,y,t=0\right) &=& 2C\sin{\left[\pi\left(r-4.5\right)\right]} \label{VIN}\, , \\
    v_{\parallel}\left(x,y,t=0\right) &=& 0 \label{VINP} \, ,
\end{eqnarray}
where $r = \sqrt{x^2 + y^2}$. This initial wave pulse can be seen in Figure \ref{SIMSNAPS} for $t=0$.

%%%%%%%%%%%%%%%%%%%%%%%%%%%%%%%%%%%%%%%%%%%%%%
\begin{figure}[t]
    \centering
    \includegraphics[width=0.5\textwidth]{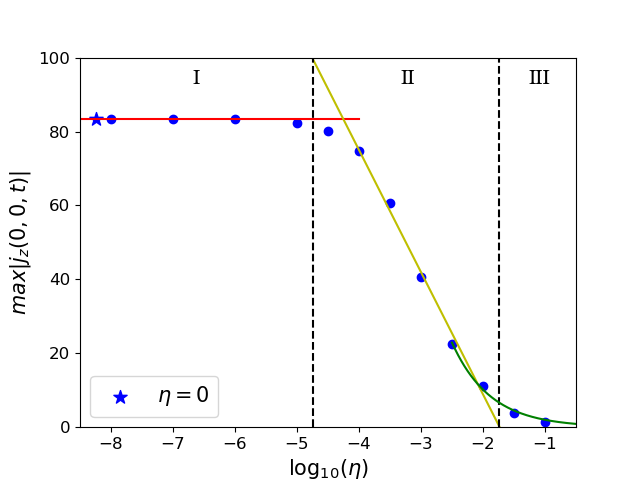}
    \caption{Plot of resistivity vs maximum amplitude of $|j_z|$, with the black dashed lines giving approximate locations for different trends and splitting the plot into three regions, labelled I, II and III respectively. The red line shows a flat trend for points where numerical resistivity dominates. The light green line gives a linear trend for intermediate resistivity values (in region II). The dark green line shows the trend for high resistivity values as the amplitude of the current approaches zero.}
    \label{JZAMP}
\end{figure}
%%%%%%%%%%%%%%%%%%%%%%%%%%%%%%%%%%%%%%%%%%%%%%%%

Due to the definition of the initial wave pulse in this way, it will result in the wave splitting into two waves -- one of these waves will move towards the null, with the other moving away --  each with an amplitude of $C$. This study is solely focused on the incoming wave, with the outgoing wave being numerically removed via a damping region which is discussed in Section \ref{BC}.

The choice for the value of constant $C$ dictates at which distance from the null the wave pulse will begin to shock. In this system, a value of $C=1$ is chosen, as per \cite{2009A&A...493..227M}, which allows for the initial wave pulse to develop a shock before it reaches the null point.

%%%%%%%%%%%%%%%%%%%%%%%%%%%%%%%%%%%%%%%%%%%%%%%%

\subsection{Boundary Conditions and Damping Region}\label{BC}

Zero gradient boundary conditions are implemented for velocity, magnetic field, density, temperature and energy. Due to the Alfv\'en speed increasing linearly with distance away from the null, waves are accelerated as they approach the boundary before reflecting. Since reflected waves would interfere with the long-term evolution at the null, the damping of these oscillations and reflections is crucial for this study.  Therefore a damping region is implemented to prevent these reflections from interfering with the region of interest around the null point.

As a result of the stretching of the grid in the numerical domain, a large damping region can be implemented due to the boundaries being positioned a large distance from the region of interest. This damping region is implemented for $r\geq 10$ where, using a combination of viscosity and artificial kinetic energy removal, oscillations and reflections that enter this region are slowly damped away, never to return to the central region of interest.

The use of viscosity as a way of damping away the outgoing oscillations (for $r\geq 10$) is implemented by introducing a viscosity term, $\nu\nabla^2 \mathbf{v}$, into the momentum equation (\ref{MOME}), where $\nu$ is the viscosity term. This term is highly spatially dependent (e.g. it is zero for $r < 10$) and is defined such that its value increases as it approaches the boundary.

%%%%%%%%%%%%%%%%%%%%%%%%%%%%%%%%%%%%%%%%%%%%%%%%
\begin{figure}[t]
    \centering
    \includegraphics[width=0.5\textwidth]{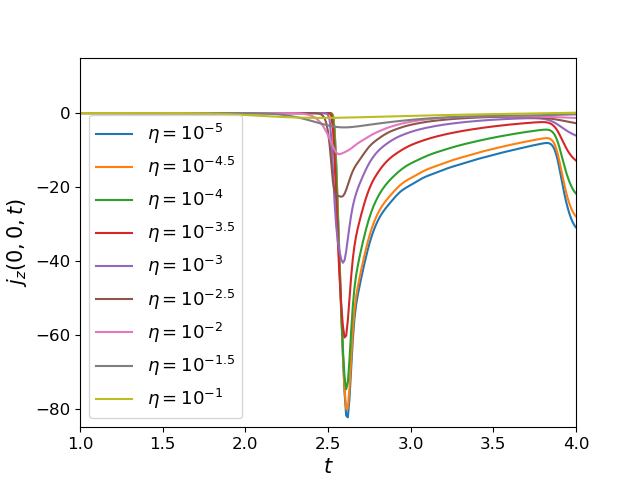}
    \caption{Overplot of $j_z(0,0,t)$ oscillations, showing effect of resistivity on amplitude of current density produced at the formation of the first current sheet. The line for each value of $\eta$ is given in the legend.}
    \label{JZOVER}
\end{figure}
%%%%%%%%%%%%%%%%%%%%%%%%%%%%%%%%%%%%%%%%%%%%%%%%

The artificial removal of kinetic energy is implemented by reducing the amplitude of the oscillations within the damping region for each iteration of the Lare2d code. This is implemented using the equation: $\mathbf{v} = \mathbf{v}/(1+a)$, where $\mathbf{v}$ is velocity and $a$ is a small amplitude defined by a number of hyperbolic tangent functions which are radially dependent. Due to the large size of the damping region, this process successfully damps away outgoing oscillations and any reflections.

It is important to note here that these damping techniques are not implemented within the uniform grid region around the null. They only occur within the damping region ($r\geq 10$).

%%%%%%%%%%%%%%%%%%%%%%%%%%%%%%%%%%%%%%%%%

\section{Results}\label{sec:results}

%%%%%%%%%%%%%%%%%%%%%%%%%%%%%%%%%%%%%%%%%%%%%%%%%%%%
%\begin{figure}[t]
%    \centering
%    \includegraphics[width=0.5\textwidth]{Fourier_Spectra_Eta=-4_bl.png}
%    \caption{Normalized Fourier spectra (in black) for $\eta=10^{-4}$ simulation with fitted Gaussian in red. The mean and standard deviation of the Gaussian fit can be found in the legend. }
%    \label{FOURIER}
%\end{figure}
%%%%%%%%%%%%%%%%%%%%%%%%%%%%%%%%%%%%%%%%%%%%%%%%%%%%

\subsection{Fast Magnetoacoustic Wave Driven Oscillatory Reconnection}

%Focusing on one simulation, which shall be referred to as the 'baseline' simulation, where the value of the resistivity is set to $\eta=10^{-4}$,

Let us now consider a simulation where the value of the resistivity is set to $\eta=10^{-4}$, which we will refer to as the ``baseline'' simulation. It is noticed that due to the definition of the initial condition, Equations (\ref{VIN}) and (\ref{VINP}), the wave pulse splits into two oppositely-propagating waves, both with amplitude $C$. As discussed previously, the outer-propagating wave will move into the damping region away from the null where it is removed from the simulation. The inwardly-propagating wave pulse moves towards the null and develops shocks on either the leading or trailing edge, depending on the region of connectivity the wave is propagating through. As these shock fronts steepen, they begin to bend the magnetic field away from the normal and eventually collapse the null point into a current sheet (Figure \ref{SIMSNAPS} for $t=2.0$ and $t=2.7$ respectively). As this current sheet is formed, jets are formed at the ends of the sheet, which can be seen in Figure \ref{SIMSNAPS} for $t=2.7$. We shall refer to this current sheet orientation, at approximately $135$\textdegree, as orientation 1.

%%%%%%%%%%%%%%%%%%%%%%%%%%%%%%%%%%%%%%%%%%%%%%%%%%%%
\begin{figure}[t]
    \centering
    \includegraphics[width=0.5\textwidth]{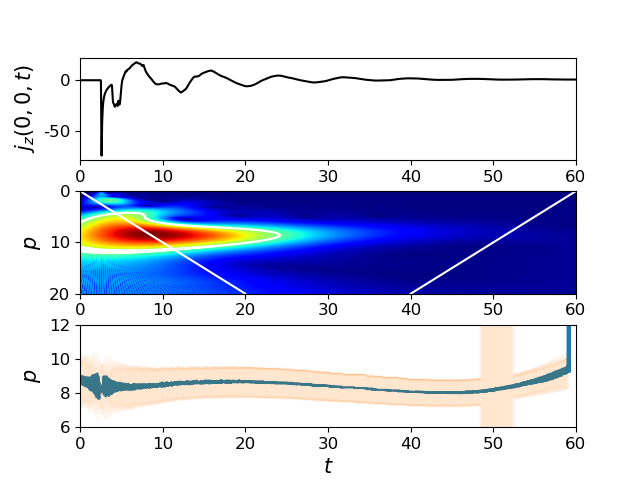} 
    \caption{Top panel shows the $j_z(0,0,t)$ oscillation for $\eta=10^{-4}$. Middle panel shows the wavelet power contour, where dark red denotes a high power and dark blue denotes low power. The slanted white lines denote the cone of influence and the white contour encompassing the main peak in the spectrum denotes the $60\%$ confidence interval. The bottom panel shows the predominant period at each time-step (dark blue) with their standard deviation (orange).}
    \label{WAVELET}
\end{figure}
%%%%%%%%%%%%%%%%%%%%%%%%%%%%%%%%%%%%%%%%%%%%%%%%%%%%

Subsequently, the length of the current sheet begins to shrink before changing orientation by $90$\textdegree \ (which we will refer to as orientation 2), as seen in Figure \ref{SIMSNAPS} for $t=6.9$. This process repeats, leading to an oscillating current sheet, which is the main characteristic of OR. \cite{2009A&A...493..227M} studied the mechanism for this oscillation in the current sheet orientation including how shocks form in the jets and affect the heating of the surrounding plasma.

This oscillation of the current sheet can be seen by calculating the $j_z$ component of the current density at the null point using Equation (\ref{VXB}). From Figure \ref{JZ} it can be seen that the current density remains at zero until a time of $t\approx 2.7$, at which point the null point collapses as the first current sheet is formed (Figure \ref{SIMSNAPS} for $t=2.7$). The value of $j_z$ drops considerably at this point, with the negative value corresponding to a current sheet of orientation 1. The value of the current then changes to become positive at a time of $t\approx 5$, which corresponds to the formation of the current sheet in orientation 2 (seen in Figure \ref{SIMSNAPS} for $t=6.9$).

%%%%%%%%%%%%%%%%%%%%%%%%%%%%%%%%%%%%%%%%%%%%
\begin{figure}[t]
    \centering
    \includegraphics[width=0.49\textwidth]{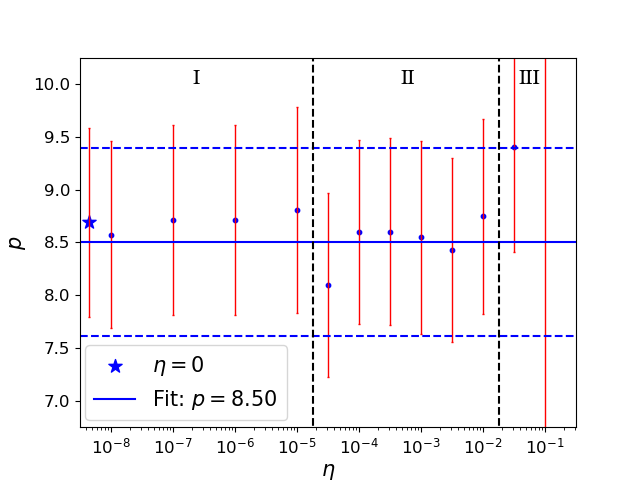}
    %\subfloat{\includegraphics[width=0.49\textwidth]{Resistivity_vs_Period_Fourier.png}}
    \caption{Resistivity vs period from wavelet analysis. The blue dots indicate the mean period values with the red bars denoting their attributed errors. The blue line indicates a mean period calculated from the period values for $\eta = 10^{-4.5}$ to $\eta = 10^{-2}$  with the blue dashed lines indicating the uncertainty on this value calculated through Propagation of Error. The blue star indicates the results from the ideal simulation. The black dashed lines split the plot into three regions for low (I), intermediate (II) and high (III) levels of resistivity as in Figure \ref{JZAMP}. }
    \label{RESPER}
\end{figure}
%%%%%%%%%%%%%%%%%%%%%%%%%%%%%%%%%%%%%%%%%%%%

It can also be seen that the oscillations in $j_z$ decay over time. This is due to the gradual shortening of the current sheets after each oscillation due to the magnetic field slowly relaxing back to the equilibrium state. It is noted here that the oscillations of $j_z$ do not oscillate around a zero value, but instead oscillate around a value of $j_z \approx 0.87$, as is shown by the blue dashed line in Figure \ref{JZ}. This is due to residual current density that remains in the system, which results from the magnetic field being very slightly non-potential at the end of the simulation (due to localised heating from the periodic, orientation-changing reconnection jets). This can be seen in Figure \ref{SIMSNAPS} for $t=60$, where there is a slight difference between the magnetic field line (black) and the $t=0$ separatrices (green).

With regards to the rest of this paper, in order to investigate the sensitivity of the system to resistivity, we perform a parameter study over eight orders of magnitude, from $\eta=10^{-8}$ to $\eta=10^{-1}$. Specifically we conduct thirteen simulations, with values $\eta=10^{-8}$, $10^{-7}$, $10^{-6}$, $10^{-5}$, $10^{-4.5}$, $10^{-4}$, $10^{-3.5}$, $10^{-3}$, $10^{-2.5}$, $10^{-2}$, $10^{-1.5}$, $10^{-1}$ as well as an ideal simulation with $\eta=0$.

%The period of the {\color{red}{OR}} system can be found by focusing on this oscillating $j_z$ signal. In the next subsections, three methods for analysing the period of the system will be discussed, along with the comparison for both the values and evolution of the period for different values of resistivity. 

%%%%%%%%%%%%%%%%%%%%%%%%%%%%%%%%%%%%%%%%%%%%

%%%%%%%%%%%%%%%%%%%%%%%%%%%%%%%%%%%%%%%%%%%%%%%%%%%%%%
\begin{center}
    \begin{table}
        \centering
        \caption{Order of fitted polynomials to period vs time plots from ANOVA results.}
        \begin{tabular}{c c}
        \hline
        Resistivity Level & Order of Fit \\
        \hline
        \hline
        $0$  &    3rd   \\
        \hline
        $10^{-5}$ &    4th   \\
        \hline
        $10^{-4.5}$ & 2nd \\
        \hline
        $10^{-4}$ &    2nd  \\
        \hline
        $10^{-3.5}$ &  4th \\
        \hline
        $10^{-3}$ &    4th   \\
        \hline
        $10^{-2.5}$ & 4th \\
        \hline
        $10^{-2}$ &    4th   \\
        \hline
        $10^{-1.5}$ & 3rd \\
        \hline
        \end{tabular}
        \label{ANOVA}
    \end{table}
\end{center}
%%%%%%%%%%%%%%%%%%%%%%%%%%%%%%%%%%%%%%%%%%%%%%%%%%%%%%

\subsection{Effect on Amplitude of $j_z$ Oscillation}\label{sec:jzamp}

%In the next few subsections, further study into how resistivity affects the OR mechanism is undertaken, focusing on the effect that resistivity has on the current density produced in the system, and in particular the $z$-component of the current density produced at the position of the null, i.e. $j_z(0,0,t)$.

Firstly, the focus will be on the amplitude of the current density produced at the null, in particular the current density produced at the formation of the first current sheet, which corresponds to the first local minimum in the current density. By comparing the values of $j_z(0,0,t)$ at the formation of the first current sheet for each of the simulations, it can be understood how resistivity affects the magnitude of the current density produced through the OR mechanism. A comparison of these values can be seen in Figure \ref{JZAMP}.

From Figure \ref{JZAMP}, it can be seen that there are three different regions of interest. The first is the region consisting of points to the left-hand-side of the plot (points for $\eta = 10^{-8}$ to $10^{-5}$). In this region, hereby referred to as region I, it is found that a change in the background resistivity has little, if any, effect on the amplitude of the maximum absolute value of the current density. We interpret this is due to the simulations not being of high enough resolution to resolve these low levels of resistivity, and we conclude that they instead are dominated by numerical resistivity. This is confirmed when comparing these values with the value from the ideal simulation ($\eta =0$). Due to this, no conclusions will be made based solely on these simulations.

Secondly, there is a region in the center of the plot for  levels of resistivity (points for $\eta = 10^{-4.5}$ to $10^{-2}$). In this region, hereby referred to as region II, we can see a clear linear trend between resistivity and the amplitude of the $|j_z(0,0,t)|$ values. 

\begin{figure}[t]
    \centering
    \includegraphics[width=0.5\textwidth]{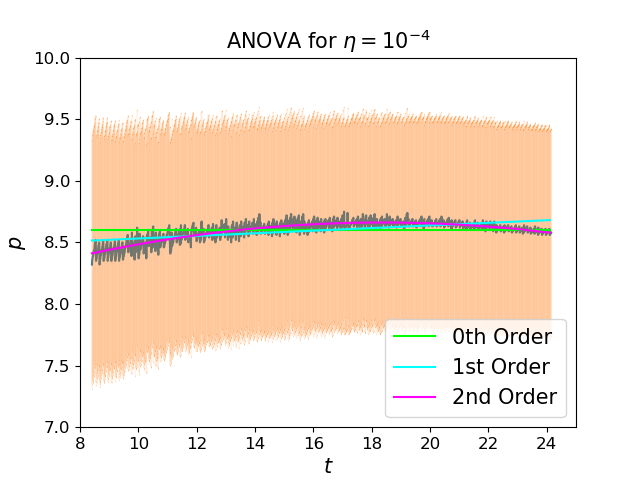}

%    \caption{Shows the analysis of variance study for our baseline simulation ($\eta=10^{-4}$). The dark blue dots show the predominant period at each time step with the purple error bars taken from the bottom panel of Figure  \ref{WAVELET}. The lime, cyan and magenta lines denote the $0^{th}$, $1^{st}$ and $2^{nd}$ order fits respectively.}
    
    \caption{The analysis of variance study for the baseline simulation ($\eta=10^{-4}$). The dark blue dots show the predominant period at each time step with the orange error bars taken from the bottom panel of Figure  \ref{WAVELET}. The lime, cyan and magenta lines denote the $0${th}, $1${st} and $2${nd} order fits respectively.}
    \label{ANOVAPLOT}
\end{figure}

The third region is that for high levels of resistivity (points for $\eta = 10^{-1.5}$ to $10^{-1}$). In this region, hereby referred to as region III, one can see the levels of \textbf{current density} flatten out as they approach zero. This trend is a trivial one as the value of $|j_z(0,0,t)|$ cannot drop below zero due to us taking the absolute value of the current.

A comparison of the first local minima can be seen in Figure \ref{JZOVER}, where the effect of resistivity on the amplitude of the $j_z(0,0,t)$ can be seen.

\subsection{Effect of Resistivity on Periodicity}\label{period}

The period of the OR system can be found by focusing on the oscillating $j_z(0,0,t)$ signal seen in Figure \ref{JZ}. In the next subsections, two methods for analyzing the period of the system are discussed, along with the comparison for both the values and evolution of the period for different values of resistivity.

\subsubsection{Wavelet Analysis}\label{Section: Wavelet Analysis}

The first method that will be used to analyze the period for the OR system is that of wavelet analysis, where for this analysis the Morlet wavelet is employed.

The results of this analysis is a contour of power against time and period (Figure \ref{WAVELET}: middle). In this contour, the dark red [blue] signifies a high [low] power. Notice also on this plot the regions for the cone of influence (COI), defined by the white slanted lines, and the 60\% confidence interval (CI), defined by the white contour encompassing the main peak.

The period of the system can be found by analyzing the results of the wavelet power contour within the region between the COI and the 60\% CI.

%%%%%%%%%%%%%%%%%%%%%%%%%%%%%%%%%%%%%%%%%%%%%%%%%%
\begin{figure}[t]
    \centering
    \includegraphics[width=0.5\textwidth]{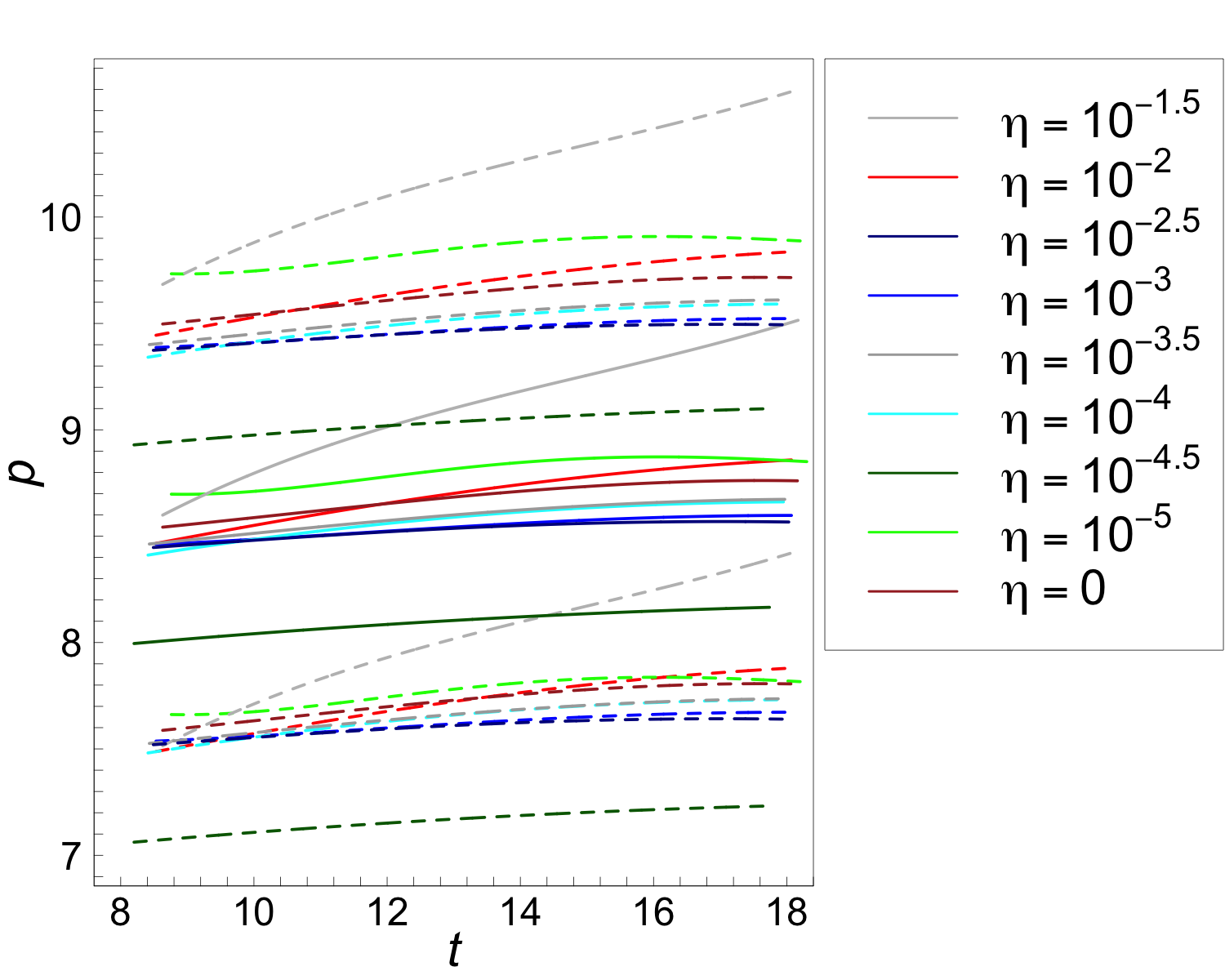}
    \caption{Prediction interval fits for simulations where $\eta=10^{-5}$ to $\eta=10^{-1.5}$ as well as the ideal case ($\eta=0$). The solid lines denote the predicted values with the dashed lines showing the attributed errors. The colors for each simulation can be found in the legend.}
    \label{PREDICT}
\end{figure}
%%%%%%%%%%%%%%%%%%%%%%%%%%%%%%%%%%%%%%%%%%%%%

This wavelet analysis begins by taking cuts of the wavelet power contour at every time step. This gives a profile of power against period for each time step. To find the predominant period at each cut a Gaussian is fitted to the profiles, with the center $\mu$ being the predominant period, and its standard deviation $\sigma$. Plotting these predominant periods (blue points) and their respective standard deviations (orange bars) against time gives the plot seen in Figure \ref{WAVELET}: bottom for $\eta =10^{-4}$. (The wavelet plots of the other simulations of varying levels of $\eta$ can be found in Appendix \ref{SUPPLOTS}, Figure \ref{WAVELETPLOTS})

From this plot, the period of the system can be calculated by taking the average value of the period within the region between the COI and the CI. The method of Propagation of Error (PoE) is used to incorporate all of the standard deviations (between the COI and CI) in Figure \ref{WAVELET}: bottom to calculate the error on this period. This process is then repeated across each of the simulations for different levels of resistivity resulting in Figure \ref{RESPER} which shows the respective periods (blue points) and attributed errors (red error bars). The black dashed lines split this plot into three regions as in Figure \ref{JZAMP}.

From Figure \ref{RESPER}, it can be seen that any change in period when resistivity is changed is negligible within the error bars. Therefore, we can fit a zeroth order trend to this plot, $p=8.50$, shown by the blue line in Figure \ref{RESPER}: left. This fit is calculated only using points from region II, $\eta=10^{-4.5}$ to $\eta=10^{-2}$, due to the size of the errors for the points where $\eta=10^{-1.5}$ and $\eta=10^{-1}$.

To calculate a resultant uncertainty of this constant period, PoE is again utilized to incorporate all the individual errors on the period values for each simulation. This uncertainty can be seen via the blue dashed lines in Figure \ref{RESPER}. These calculations yield a value for the period of $p_{\text{wavelet}}=8.50 \pm 0.89$, irrespective of the value for the resistivity.

To corroborate these results, Fourier transforms of the $j_z(0,0,t)$ profiles were calculated with the results found in Appendix \ref{Fourier}.

%%%%%%%%%%%%%%%%%%%%%%%%%%%%%%

%%%%%%%%%%%%%%%%%%%%%%%%%%%%%%

%%%%%%%%%%%%%%%%%%%%%%%%%%%%%%%%%%%%%%%%%%%%%

\subsubsection{Analysis of Variance and Prediction Intervals}

Section \ref{Section: Wavelet Analysis} yields values for the period with the assumption that this period is constant throughout the duration of the simulations. However, in Figure \ref{WAVELET}: bottom, there appears to be a change in the period with respect to time. To test the significance of this change, Analysis of Variance (ANOVA) tests \citep[]{scheffe1999analysis}{} can be carried out by testing the significance of higher-order terms in a polynomial fit. This results in the lowest-order polynomial which accurately fits to the data to be found. A visual representation of this ANOVA test on the baseline simulation can be seen in Figure \ref{ANOVAPLOT}.

This technique is carried out on the plots of period against time in Figure \ref{WAVELET}: bottom for each of our simulations, with the resulting order of polynomial fit for each simulation given in Table \ref{ANOVA}. Due to the high levels of decay for the $\eta = 10^{-1}$ simulation, we do not include this simulation in this analysis as the errors from the wavelet power contour are too large. Due to the dominance of numerical resistivity for simulations within region I, these simulations are also omitted from this analysis.

It can be noticed from Table \ref{ANOVA} that despite changing the level of resistivity across multiple orders of magnitude, the evolution of the periods of each simulation can be described by similarly ordered polynomials. As a further test of these fits, the prediction intervals of the fitted lines against the data are calculated.

To incorporate the error bars calculated in Figure \ref{WAVELET}: bottom, the Von Neumann rejection method is used to produce a Gaussian distribution of 1000 random points at each time step. These Gaussian distributions take the values for parameters $\mu$ and $\sigma$, directly from the values in the bottom panel of Figure \ref{WAVELET} (i.e. Section \ref{Section: Wavelet Analysis}). The prediction intervals are then calculated using these distributions.

%%%%%%%%%%%%%%%%%%%%%%%%%%%%%%%%%%%%%%%%%%%%%%%%
\begin{figure}[t]
    \centering
    \includegraphics[width=0.49\textwidth]{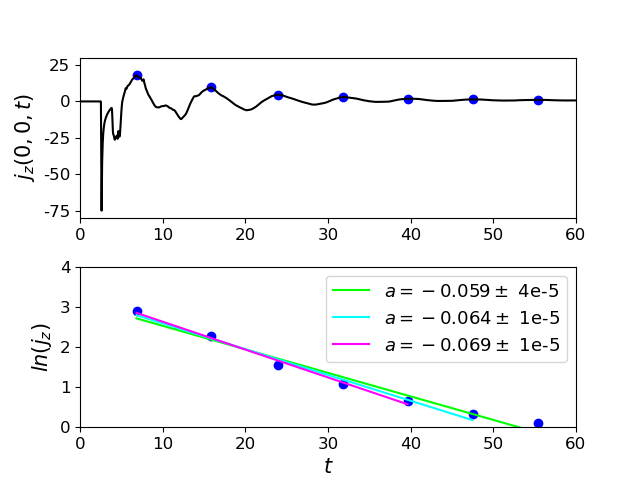}
    \caption{Calculation of the decay rate of the $j_z(0,0,t)$ oscillation when $\eta = 10^{-4}$. In the top row, the blue points correspond to the location of the local maxima of the $j_z(0,0,t)$ profile (solid black line). The bottom row shows the values of the local maxima but on a natural logarithmic scale (blue points) with a linear fit to these values for 5, 6, and 7 local maxima shown by the magenta, cyan and lime lines respectively.. The legend shows the fitted decay constant of the linear fit with the attributed error.}
    \label{DECAY}
\end{figure}
%%%%%%%%%%%%%%%%%%%%%%%%%%%%%%%%%%%%%%%%%%%%%%%%

The prediction interval was calculated using the in-built function ``Predict'' in R, with the underlying statistics calculated by the equation:
\begin{equation*}
    y_{p_{\pm}} = \hat{y}_h \pm t_{a/2,n-2}\sqrt{MSE\left(1 + \frac{1}{n} + \frac{\left(x_k - \Bar{x}\right)^2}{\sum\left(x_i - \Bar{x}\right)^2}\right)}\, ,
\end{equation*}
\\
where $y_{p_{\pm}}$ is the prediction interval, $\hat{y}_h$ is the fitted data, $n$ is the number of data points, $MSE$ is the mean square error, $x_k$ is the predictor value, $\bar{x}$ is the mean value and $t_{a/2,n-2}$ is the critical $t$-value where $a$ is the confidence level with $n-2$ degrees of freedom. In this study we take the $95\%$ prediction interval.

Figure \ref{PREDICT} shows the prediction intervals for each simulation. In this plot, the range of time is fixed to that of the shortest range between the COI and 60\% CI (Figure \ref{WAVELET}: middle) across the simulations. This is done so that a direct comparison of the evolution for the period of each simulation can be made.

It can be seen from Figure \ref{PREDICT} that the fits for each simulation (solid lines) lie within the prediction interval of each simulation (dashed lines). This suggests that the variation in the fits across the simulations is negligible in relation with the attributed errors. Due to this it can be stated that none of the models are significantly different from one another.

%%%%%%%%%%%%%%%%%%%%%%%%%%%%%%%%%%%%%

%%%%%%%%%%%%%%%%%%%%%%%%%%%%%%%%%%%%%%%%%%%%%%%%%%

\subsection{Effect on Decay Rate of $j_z(0,0,t)$ Oscillation} \label{sec:decay}

%%%%%%%%%%%%%%%%%%%%%%%%%%%%%%%%%%%%%%%
\begin{figure}[t]
    \centering
    \includegraphics[width=0.49\textwidth]{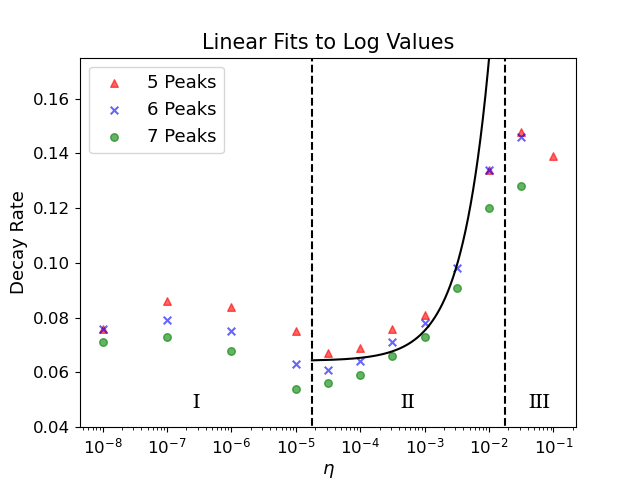}

    \caption{Resistivity vs. decay rate taken from linear fit to log-value plots (Fig. \ref{DECAY} bottom row). The plot is split into three regions (I, II and III), as in Figure \ref{JZAMP}, shown by the black dashed lines. Each simulation has multiple points depending on how many local maxima were used when fitting the data, indicated in the legend. The black solid line is an exponential fit, fitted to the points in the intermediate resistivity range (region II) for the decay rates calculated using 6 local maxima.}
    \label{DECAYCOMP}
\end{figure}
%%%%%%%%%%%%%%%%%%%%%%%%%%%%%%%%%%%%%%%

A further study can be made into how resistivity affects the decay rate of the $j_z$ oscillation at the null. This is done using a similar method to that used in \cite{2022ApJ...933..142K}, namely considering the local maxima of the oscillations in $j_z(0,0,t)$ and finding the rate at which these maxima decay throughout the oscillation.

To find the decay rate of these maxima, the natural logarithm ($\ln$) is taken of these local maxima and then a linear trend fitted, where the linear trend is simply defined as $y_{\text{lin}} = at + b$, where $a$ is a decay constant and $b$ is a constant. This fit can be seen in Figure \ref{DECAY}: bottom row, where  the values of the decay constants can be found in the legend with their corresponding errors.

This fitting is repeated for each of the simulations, where for each simulation, the number of local maxima to which the fit is made is varied, to investigate the effect that this has on the decay rate values calculated. The plots of the decay rates for each simulation can be found in Appendix \ref{SUPPLOTS}, Figure \ref{DECAYPLOTS}. The decay rates for each simulation can be seen in Figure \ref{DECAYCOMP}. In this plot, decay rate values calculated from $5$, $6$ and $7$ peaks are indicated by the triangle, cross and circle markers respectively, with the same separating lines (black dashed) as in Figure \ref{JZAMP} used, with the resistivity regions labeled in the same fashion.

%%%%%%%%%%%%%%%%%%%%%%%%%%%%%%%%%%

%%%%%%%%%%%%%%%%%%%%%%%%%%%%%%%%%%%%

\subsection{Effect on Ohmic Heating}\label{sec:ohm}

One effect that the generation of current density can have on a system is heating, in particular ohmic heating, which is directly related to the currents produced and the levels of resistivity in the system, with the relation defined by $\eta j^2$.

\begin{figure}
    \centering
    \includegraphics[width=0.5\textwidth]{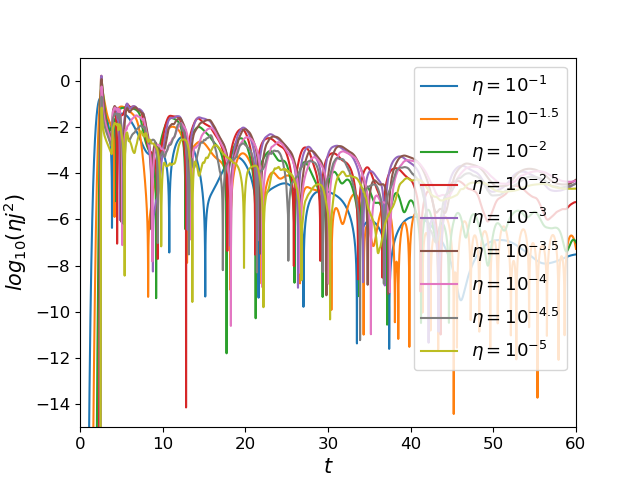}
    \caption{Time evolution of ohmic heating at the null for different levels of resistivity, presented as $\log_{10}{( \eta j_z ^2 )}$. The lines for each simulation can be found using the legend.}
    \label{OHMICCOMP}
\end{figure}

Figure \ref{OHMICCOMP} shows the evolution of ohmic heating at the null for each simulation. It can be seen from this plot, due to direct correlation with the oscillations in $j_z(0,0,t)$, that the level of ohmic heating also oscillates with the formation of each subsequent current sheet. The lines for $\eta=10^{-8}, \, 10^{-7}$ and $10^{-6}$ are not included in this figure due to them lying on top of the line of $\eta=10^{-5}$, however their points are included in Figure \ref{MAXOHM}.

For a direct comparison, the maximum value of the ohmic heating for each simulation is taken and compared. Just like for the amplitude of $j_z(0,0,t)$ study previously, the sampling of the maximum value corresponds to the formation of the first current sheet. The comparison for this maximum value can be seen in Figure \ref{MAXOHM}.

%%%%%%%%%%%%%%%%%%%%%%%%%%%%%%%%%
%%%%%%%%%%%%%%%%%%%%%%%%%%%%%%%%%%%%%%
%%%%%%%%%%%%%%%%%%%%%%%%%%%%%%%%%%%%%%%%%%
\section{Conclusions}\label{sec:conclusions}

In this paper a study into the OR mechanism in a 2D, initially cold plasma is made, developing on previous works by investigating the effect of resistivity on the system.

By solving the resistive MHD equations, using the Lare2d code, the level of background resistivity is varied across eight orders of magnitude ($\eta=10^{-8}$ to $\eta=10^{-1}$, with also an ideal MHD case where $\eta = 0$), allowing a study of how this variation affects key characteristics of the OR mechanism such as the current density produced, ohmic heating and period of the oscillating system. The choice of initial condition in this paper consists of a circular wave pulse surrounding a null point, such as in previous works by \cite{2009A&A...493..227M} and \cite{2022ApJ...925..195K, 2023ApJ...943..131K}, which initiates OR.

%%%%%%%%%%%%%%%%%%%%%
%%% I would steer clear of 'collapses the null'
%%%%%%%%%%%%%%%%%%%%

%as it collapses the null.

%For this parameter study the level of background resistivity is varied (i.e. $\eta = 10^{-8}$ to $10^{-1}$ with also an ideal MHD case where $\eta = 0$).

%%%%%%%%%%%%%%%%%%%%%%%%%%%%%%%

\begin{figure}
    \centering
    \includegraphics[width=0.5\textwidth]{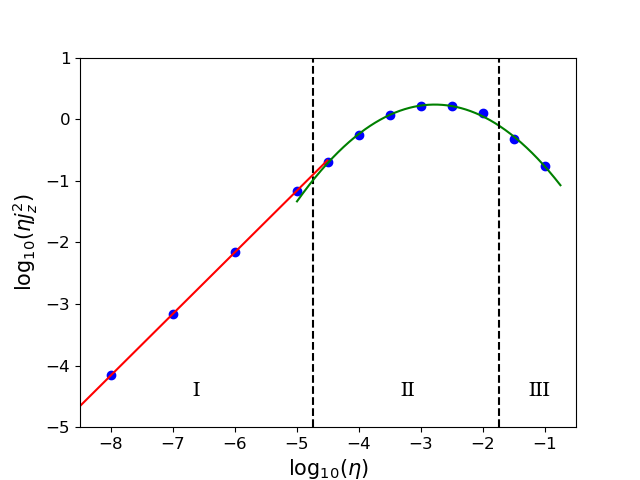}
    \caption{Resistivity vs. maximum ohmic heating. The black dashed lines split the plot into three regions for low (I), intermediate (II) and high (III) levels of resistivity as in Figure \ref{JZAMP}. The solid red line shows a linear trend, fitted to the values for low resistivity where numerical resistivity dominates. The green line shows a quadratic trend for the intermediate and higher values of resistivity.}
    \label{MAXOHM}
\end{figure}

%%%%%%%%%%%%%%%%%%%%%%%%%

For each of these simulations, the current density profile at the location of the null point, $j_z(0,0,t)$, were calculated with the maximum absolute value of the amplitudes of the current oscillations compared. It was found that for ``low'' levels of resistivity (region I of Figure \ref{JZAMP}), there was little or no change in the amplitude of $|j_z(0,0,t)|$. We explain this by concluding that this is due to the numerical resistivity of the simulation dominating the system in this region. This can be seen by the straight line plotted through these points. This interpretation is also corroborated by the good correspondence between this line and the value for $\eta = 0$, where the resistivity in the ideal simulation is of course solely numerical. Due to this, no conclusions can be made from these simulations.

It is found that for `high' levels of resistivity, region III in Figure \ref{JZAMP}, the level of the current flattens out as the absolute value approaches zero. This is shown by the green, decaying exponential fit in Figure \ref{JZAMP}.

In region II, the behavior is not affected by the numerical resistivity, nor is it affected by the current levels nearing zero. It is found that there is a direct relationship between resistivity and the amplitude of the currents produced in the system. This is indicated by the linear fit to the points in region II (light green line in Figure \ref{JZAMP}).

From the current density profiles, the period of the oscillations within the profiles were calculated using wavelet analysis. It was found that there is no significant change in the period when resistivity is increased across multiple orders of magnitude, suggesting an independence between resistivity and the periodicity of OR for our choice of parameters and assumptions. It is found that there is a good correspondence between the constant period values calculated from the wavelet analysis and Fourier transform, i.e. $p_{\text{wavelet}}=8.50 \pm 0.89$ and $p_{\text{fourier}}=8.59 \pm 1.50 $.

These values for the period are in computational units due to the system being non-dimensionalized. However, one can re-dimensionalize these values by multiplying them by the dimensional constant $t_0 = L/v_0$, where $v_0 = B_0/\sqrt{\mu_0\rho_0}$. Taking values $L = 1$ Mm, $B_0 = 1.4 $ G and $\rho_0 = 10^{-12}$ kg m$^{-3}$ for the characteristic length, magnetic field strength and density respectively, we find $t_0 = 7.78$ s. This value gives a period for our simulations of $p_{\text{wavelet}} = 66.13 \pm 6.92$ s and $p_{\text{fourier}} = 66.83 \pm 11.67$ s, calculated using wavelet analysis and Fourier transforms respectively. However, it is important to note that Equation (\ref{MAGEQ}) is scale free. Therefore there is freedom in the choice of values when re-dimensionalizing.

In addition, a statistical approach was also used through ANOVA and a prediction interval calculation to study the evolution of the period in the current density profiles from the wavelet analysis. It was found that the evolution of the period for each simulation can be described by similarly ordered fits. When the prediction intervals for each simulation were calculated, based on these fits, it was found that each evolution lies within the prediction interval for all other fits. Therefore we conclude that these fits are not significantly different from the other. This again suggests that there is an independence between resistivity and periodicity.

We conclude that there is an independence between resistivity and the periodicity of OR. 

This result of an independence between resistivity and periodicity of OR has to be discussed in the context of previous studies, specifically that of \cite{1991ApJ...371L..41C}. This result seems to contradict the result of their paper. In their study they found a direct link between resistivity and period of the oscillating current sheet. They found the frequency of the oscillation for the fundamental mode to be defined by $\approx 2\ln S $ where $S = 1/\eta$ is the Lundquist number. However, it is important to note that in \cite{1991ApJ...371L..41C}, a linearized system within a closed cylindrical domain was solved, whereas in the study presented in this paper the OR mechanism is investigated using the nonlinear MHD equations with effective open boundaries. Thus we conclude that the study presented here and that of \cite{1991ApJ...371L..41C} are investigating fundamentally different systems.

From the current density profiles for each of the simulations, an investigation was made into how resistivity affects the decay rate for the oscillations of $j_z(0,0,t)$. By finding the local maxima within the profiles, the decay rate was calculated by taking the natural logarithm (ln) of the maxima and making a linear fit to those points. In Figure \ref{DECAYCOMP}, it was found that the decay rate increases as the level of resistivity increases, with the exact nature of this increase being defined by an exponential function (solid black line). In the region where numerical resistivity dominates (region I), no fit was made. When calculating the relationship between resistivity and decay rates, only the values in region II were fitted. It was found that the decay rates calculated for `high' levels of resistivity in region III flatten out suggesting a possible maximum decay rate for the oscillations.

%%%%%%%%%%%%%%%%%%%%%%%%%%%%%%%%%%%%%%%%%%%%%%%%%%%%%

The ohmic heating produced at the null point was calculated, with the oscillations in the levels of ohmic heating for each of the simulations being directly linked to the oscillations in the current density (Figure \ref{OHMICCOMP}). For a more direct comparison, the maximum value of this ohmic heating for each of our simulations was taken. 

Similar to previous sections, Figure \ref{MAXOHM} is split into regions I, II and III. Region I, where numerical resistivity dominates, demonstrates a linear increase in ohmic heating as resistivity increases. This linear trend is due to fact that, as seen in the previous sections, we see little change in the levels of current density in the system for different choices of $\eta$. Therefore the increase that we see in the ohmic heating is solely due to the increase in the level of the background resistivity. 

In region II, a distinct quadratic trend between ohmic heating and resistivity is found. This trends suggests that there may be an ``optimal'' value of resistivity in terms of maximizing ohmic heating. 

This quadratic trend also extends to region III, and follows from the study on the amplitude of the current (Section \ref{sec:jzamp}), where for high resistivity levels we see a high level of damping on $j_z(0,0,t)$ (Figure \ref{JZAMP}). This strong effect on the current in turn will have an effect on the ohmic heating, which can be seen in Figure \ref{MAXOHM}.

To summarise, this study shows that there is an independence between background resistivity and the periodicity of the oscillatory reconnection mechanism for the parameter values chosen. However, it is also found that for values above the level of numerical resistivity, background resistivity has a distinct effect on the amplitude and decay rate of the current density oscillations at the null, as well as the levels of ohmic heating.

%%%%%%%% 
% the paragraph below is unneccessary, since you've covered this already
%%%%%%%%%%%%%%%%%%%

%To summarise, a study into how resistivity affects the oscillatory reconnection system has been conducted and found, as the main result, an independence between resistivity and the periodicity of oscillatory reconnection. This result was confirmed by using three techniques. It is also found, however, that resistivity does effect other parameters in the system, such as current amplitude, oscillation decay rate and Ohmic heating levels

%%%%%%%%%%%%%%%%%%%%%%%%%%%%%%

\section*{Acknowledgments}
All authors acknowledge the UK Science and Technology Facilities Council (STFC) for support from grant No. ST/X001008/1. This work used the Oswald High Performance Computing facility operated by Northumbria University (UK). The data that support the findings of this study are available from the corresponding author upon reasonable request. All authors also thank Natasha Jeffrey for her helpful insight during this study.

%%%%%%%%%%%%%%%%%%%%%%%%%%%%%%%%%%%%%%

\appendix

\section{Fourier Transform}\label{Fourier}
\begin{figure}[h]
    \centering
    \includegraphics[width=0.49\textwidth]{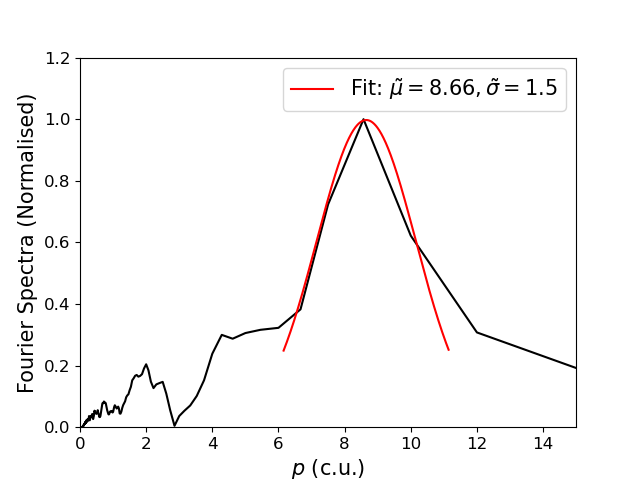}
    \includegraphics[width=0.49\textwidth]{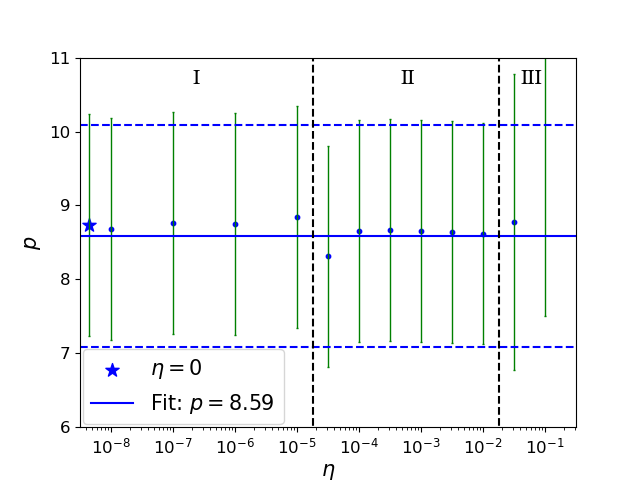}
    \caption{Left: Fourier spectra (in black) for $\eta=10^{-4}$ simulation, which is normalized to the maximum power, with fitted Gaussian in red. The mean and standard deviation of the Gaussian fit can be found in the legend. Right: Resistivity vs period from Fourier spectra. The blue points indicate the mean periods taken from the Gaussian fits for each Fourier spectrum, with the green error bars being taken from the standard deviation. The blue line indicates a mean period calculated from the period values for $\eta = 10^{-4.5}$ to $\eta = 10^{-2}$  with the blue dashed lines indicating the uncertainty on this value calculated through Propagation of Error. The blue star indicates the results from the ideal simulation. The black dashed lines split the plot into three regions for low (I), intermediate (II) and high (III) levels of resistivity as in Figure \ref{JZAMP}.}
    \label{FOURIERAPP}
\end{figure}

To corroborate the results of Section \ref{Section: Wavelet Analysis}, a second technique is used to calculate the period of each simulation: that of the Fourier transform, which outputs a spectra for periods within the system. Figure \ref{FOURIERAPP}: left shows the Fourier spectra for the baseline simulation ($\eta=10^{-4}$) which has been normalized by the maximum power of the spectra, from which a Gaussian has been fitted to the main ``peak" of each spectrum, taking the mean, $\tilde{\mu}$, of this fit as the predominant period of the system and the standard deviation, $\tilde{\sigma}$, as the attributed error to this value. The values for all simulations can be found in Appendix \ref{SUPPLOTS},  Figure \ref{FOURIERPLOTS}.

Repeating this process for all of the simulations results in Figure \ref{FOURIERAPP}: right, which shows the period (blue dots) and the respective errors (green error bars) against resistivity. Due to the size of these errors a zeroth order trend line can be fitted to this plot, shown by the blue line. This fit is only calculated using periods for $\eta=10^{-4.5}$ to $\eta=10^{-2}$ due to the size of the errors for $\eta=10^{-1.5}$ and $\eta=10^{-1}$. 

Similar to Section \ref{Section: Wavelet Analysis}, the uncertainty of this fit is calculated using PoE to incorporate the error bars in Figure \ref{FOURIERAPP}: right, and are shown by the blue dashed lines. This results in a value for the period of $p_{\text{fourier}}=8.59 \pm 1.50$, which shows a good agreement with the period calculated from the wavelet analysis.

\section{Supplemental Plots}\label{SUPPLOTS}

\begin{figure}
    \centering
    \includegraphics[width=2.25in]{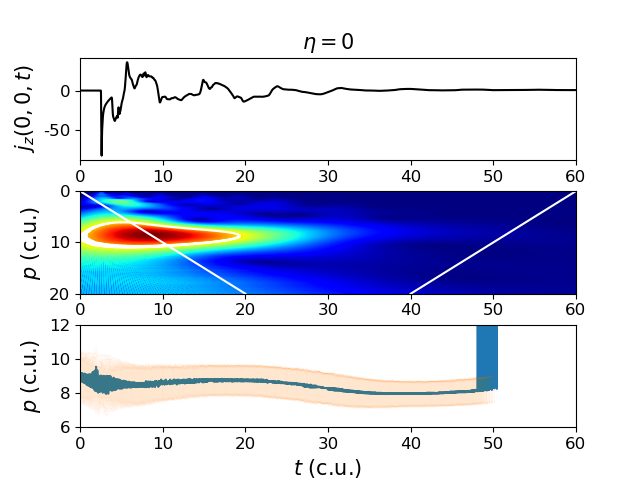}
    \includegraphics[width=2.25in]{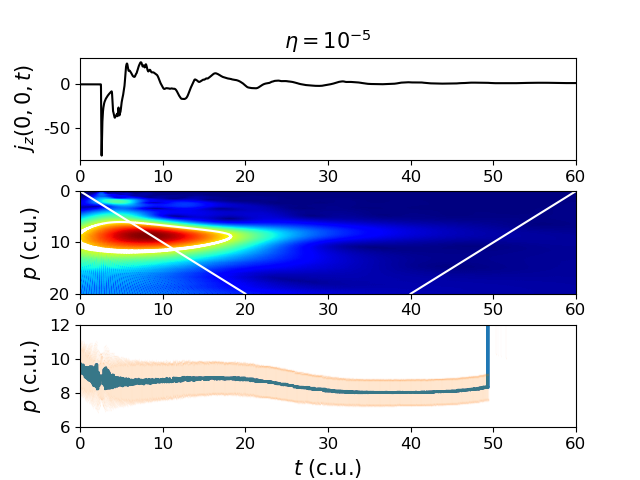}
    \includegraphics[width=2.25in]{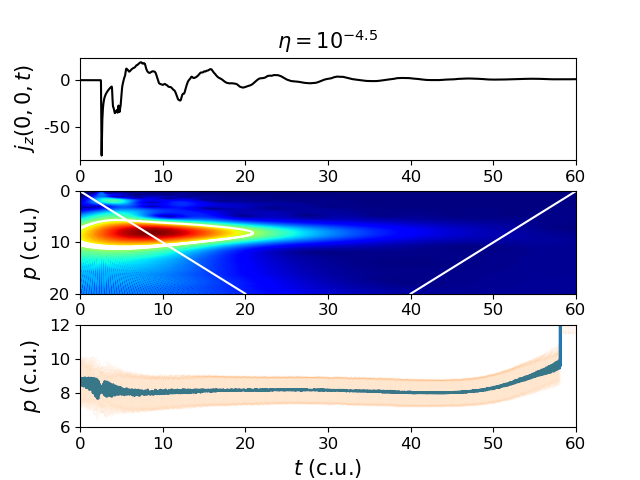} \\
    \includegraphics[width=2.25in]{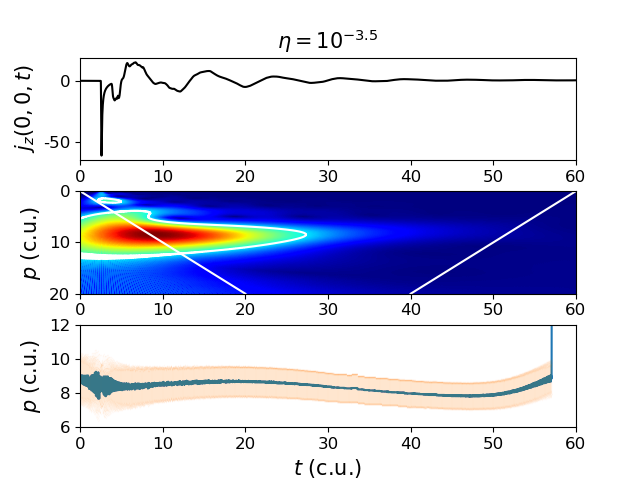} 
    \includegraphics[width=2.25in]{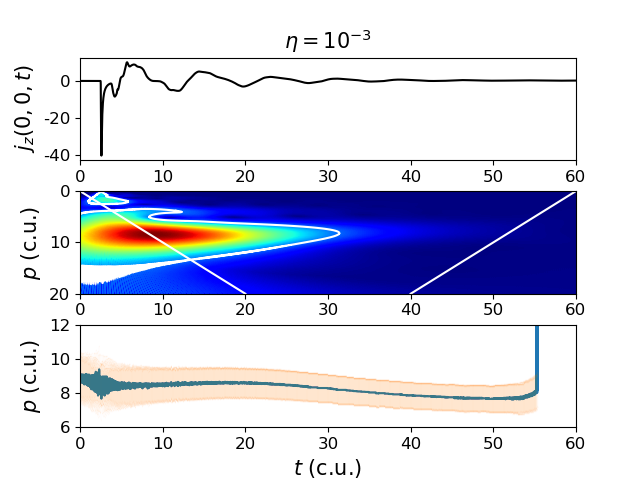}
    \includegraphics[width=2.25in]{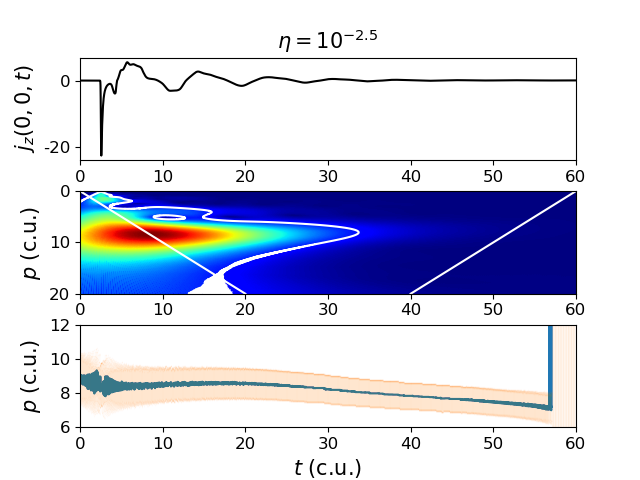} \\
    \includegraphics[width=2.25in]{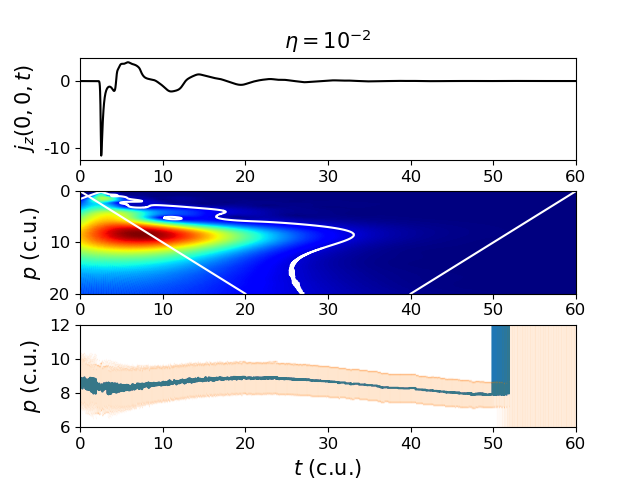} 
    \includegraphics[width=2.25in]{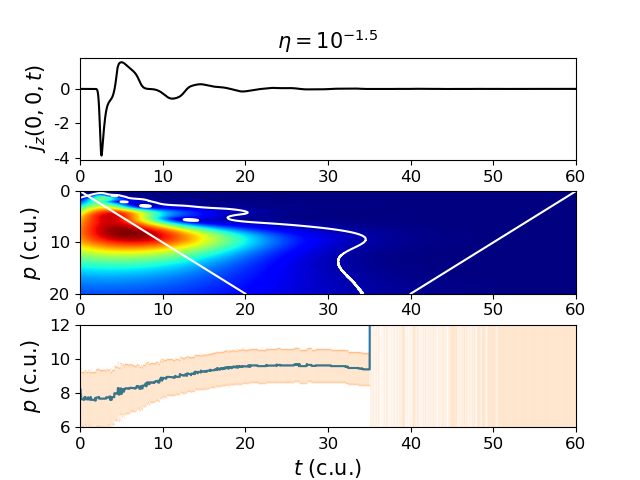}
    \includegraphics[width=2.25in]{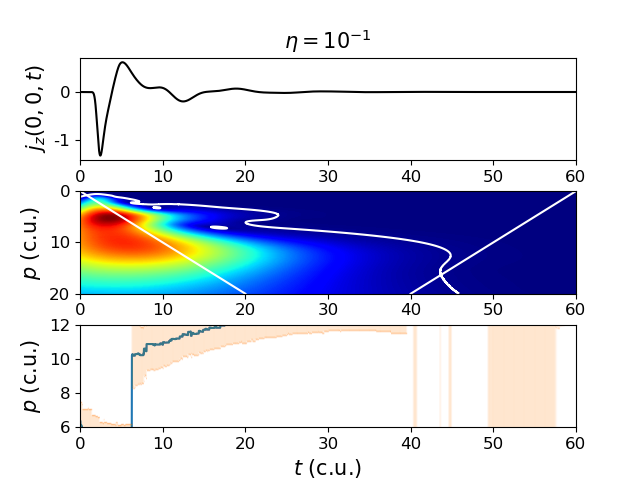}
    \caption{Panels showing wavelet analysis of current density signals, $j_z(0,0,t)$ at the null for increasing levels of background resistivity, from $\eta=0$ and $\eta=10^{-7}$ to $\eta=10^{-1}$. Results for $\eta=10^{-8},\, 10^{-7}$ and $10^{-6}$ are not shown due to their similarity with $\eta=0$ and $10^{-5}$. Similarly, $\eta=10^{-4}$ is not shown due to being available in Figure \ref{WAVELET}. For each panel, the top panel shows the  evolution of $j_z(0,0,t)$. Middle panel shows the wavelet power contour, where dark red denotes a high power and dark blue denotes low power. The white lines spanning from $(t=0,p=0)$ to $(t=20,p=20)$ and $(t=40,p=20)$ to $(t=60,p=0)$ denote the cone of influence and the white contour encompassing the main peak in the spectrum denotes the $60\%$ confidence interval. The bottom panel shows the predominant period at each time-step (dark blue) with their attributed standard deviation (orange).}
    \label{WAVELETPLOTS}
\end{figure}

\begin{figure}
    \centering
    \includegraphics[width=2.25in]{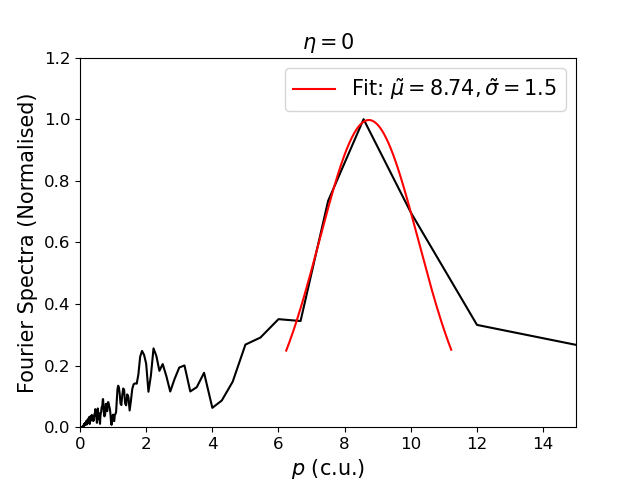}
    \includegraphics[width=2.25in]{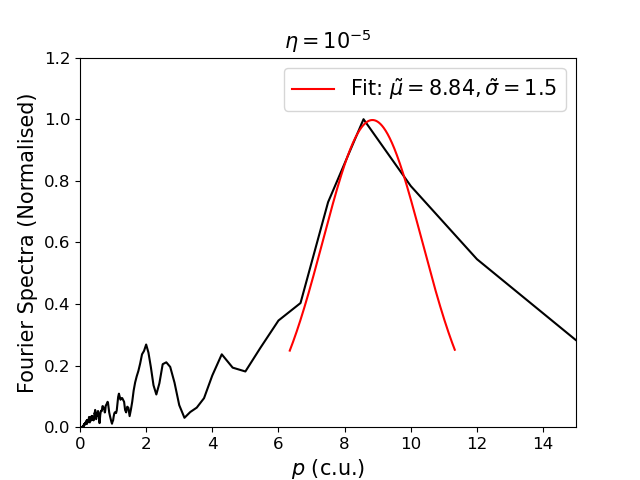} 
    \includegraphics[width=2.25in]{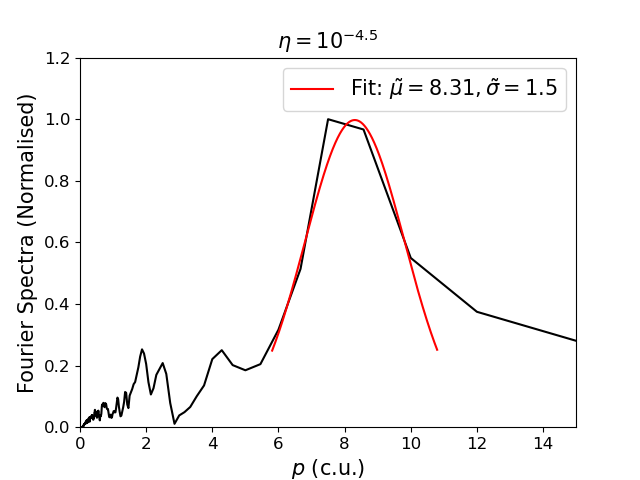} \\
    \includegraphics[width=2.25in]{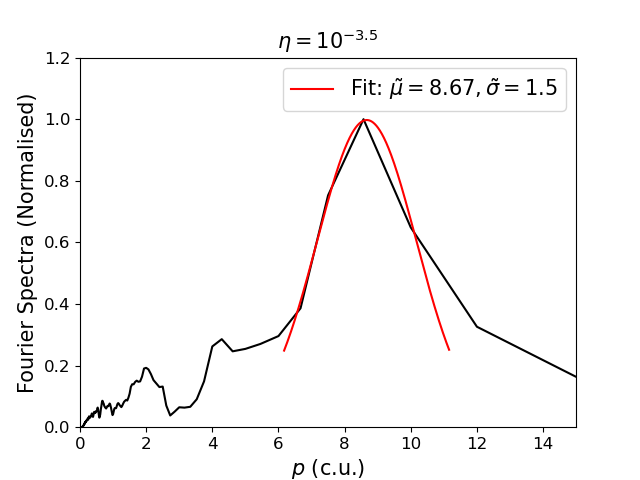} 
    \includegraphics[width=2.25in]{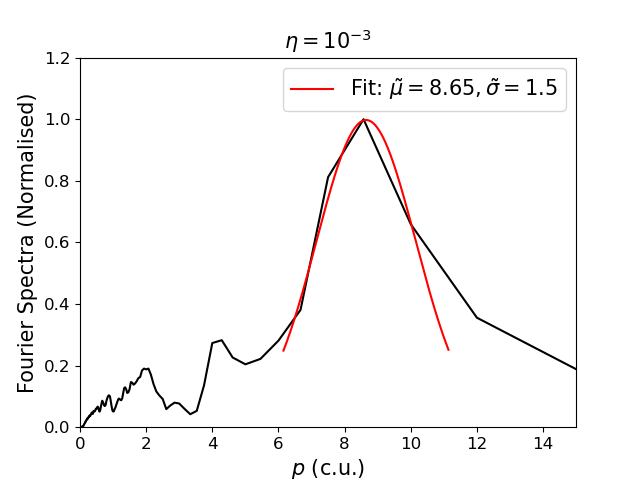}
    \includegraphics[width=2.25in]{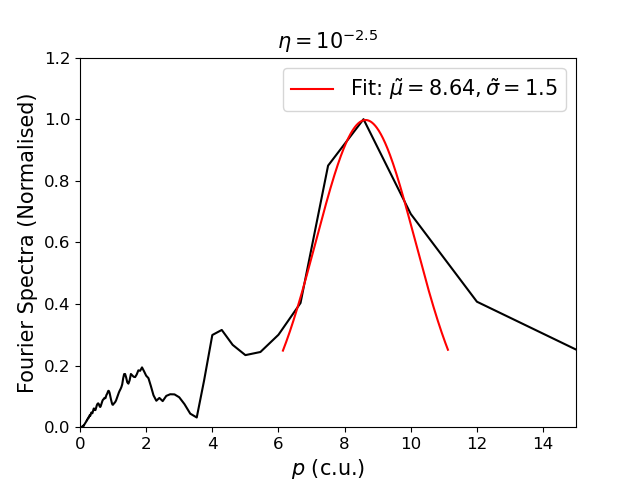} \\
    \includegraphics[width=2.25in]{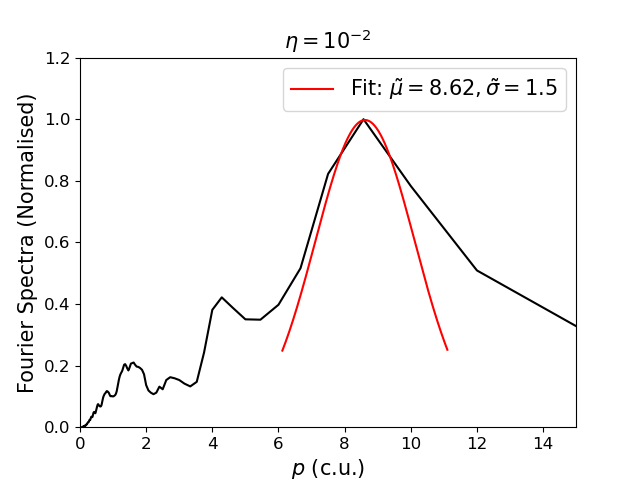}
    \includegraphics[width=2.25in]{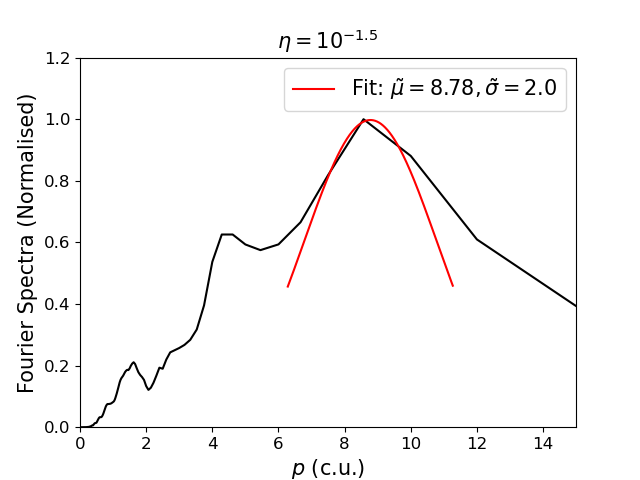}
    \includegraphics[width=2.25in]{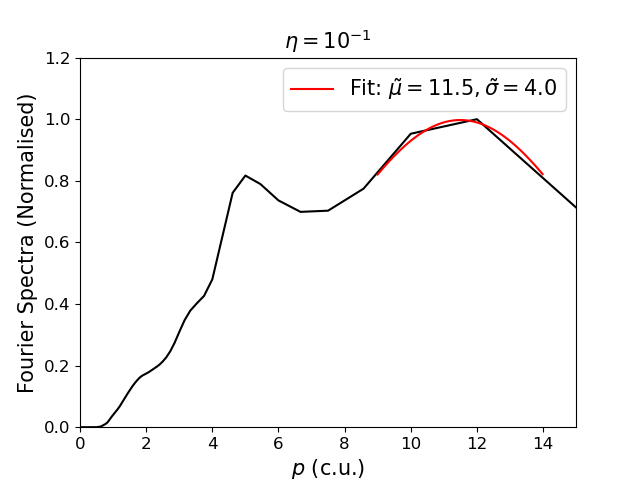}
    \caption{Panels showing the Fourier spectra for each simulation (black line) that is normalized to the maximum power, with a fitted Gaussian (red line) to the main peak of the spectra, for increasing levels of background resistivity, from $\eta=0$ and $\eta=10^{-5}$ to $\eta=10^{-1}$. Results for $\eta=10^{-8},\, 10^{-7}$ and $10^{-6}$ are not shown due to their similarity with $\eta=0$ and $10^{-5}$. Similarly, $\eta=10^{-4}$ is not shown due to being available in Figure \ref{FOURIERAPP}. The mean, $\tilde{\mu}$, and standard deviation, $\tilde{\sigma}$, are given in the legend for each panel.}
    \label{FOURIERPLOTS}
\end{figure}

\begin{figure}
    \centering
    \includegraphics[width=2.25in]{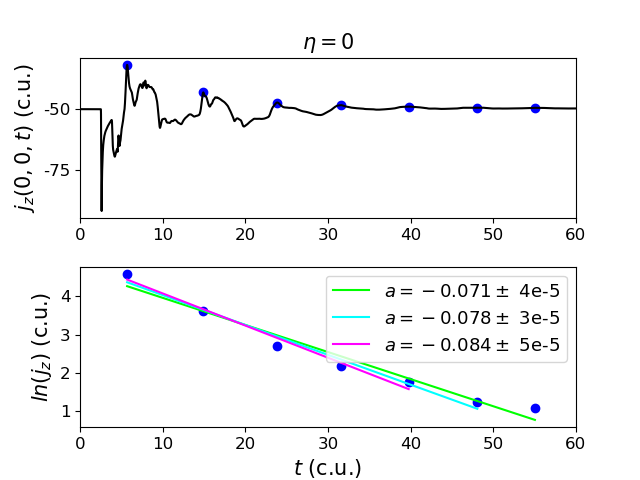}
    \includegraphics[width=2.25in]{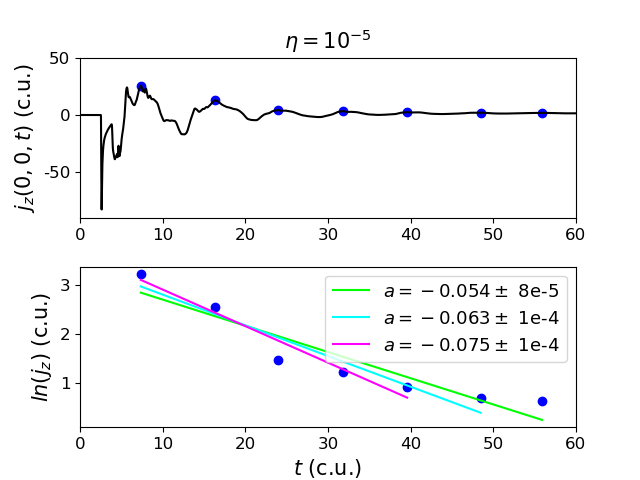} 
    \includegraphics[width=2.25in]{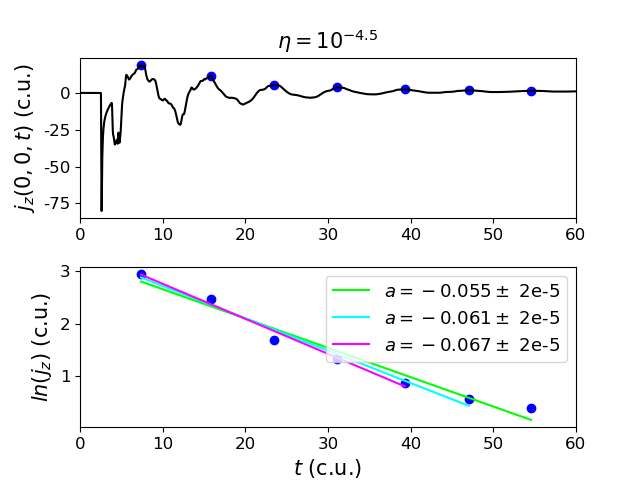} \\
    \includegraphics[width=2.25in]{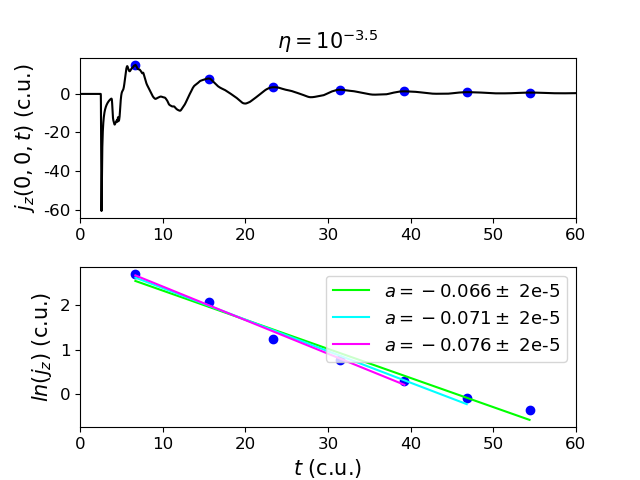} 
    \includegraphics[width=2.25in]{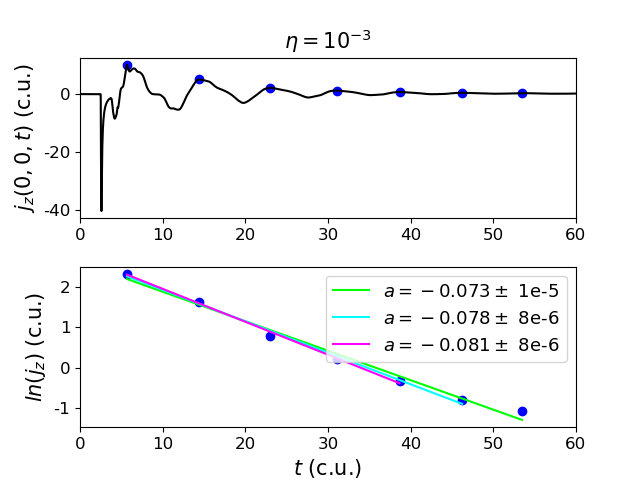}
    \includegraphics[width=2.25in]{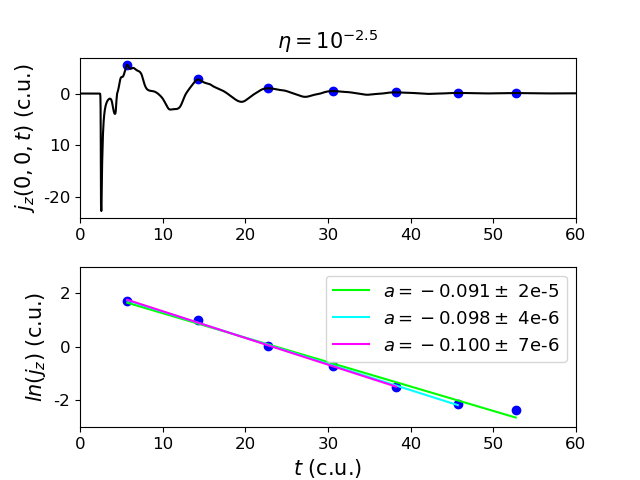} \\
    \includegraphics[width=2.25in]{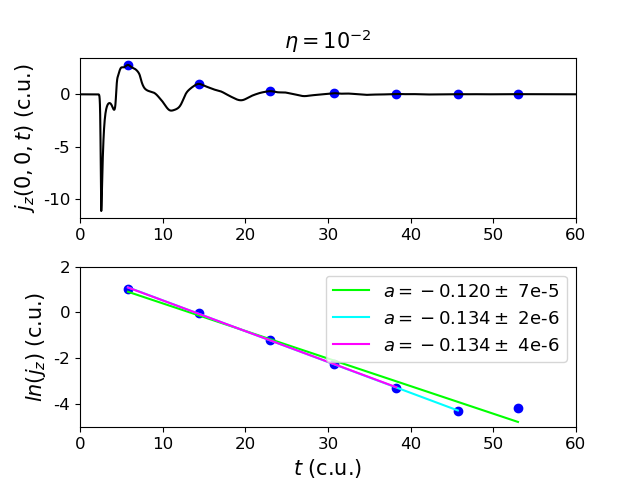}
    \includegraphics[width=2.25in]{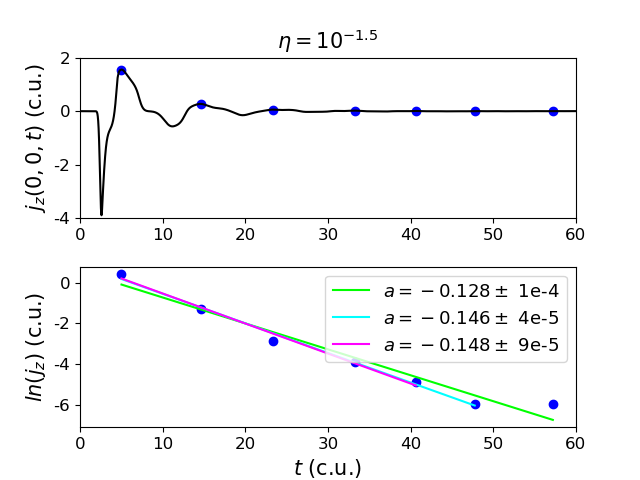}
    \includegraphics[width=2.25in]{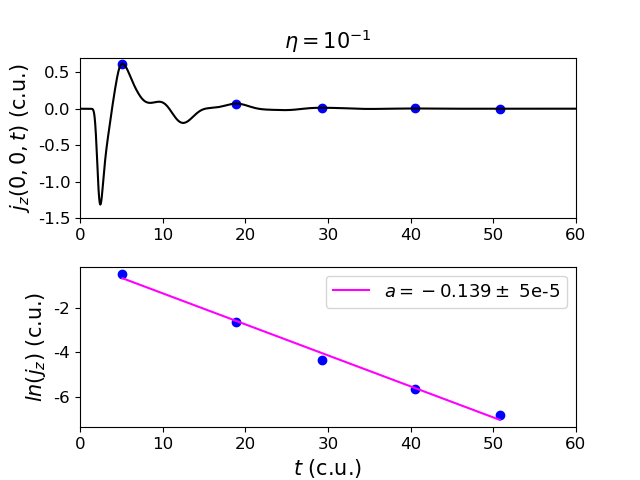}
    \caption{Calculation of the decay rate of the $j_z(0,0,t)$ oscillation for   increasing levels of background resistivity, from $\eta=0$ and $\eta=10^{-5}$ to $\eta=10^{-1}$. Results for $\eta=10^{-8},\, 10^{-7}$ and $10^{-6}$ are not shown due to their similarity with $\eta=0$ and $10^{-5}$. Similarly, $\eta=10^{-4}$ is not shown due to being available in Figure \ref{DECAY}. In the top panels, the blue points correspond to the location of the local maxima of the $j_z(0,0,t)$ profile (solid black line). The bottom panels show the natural logarithm values of the local maxima (blue points) with linear fits to these values with the magenta, cyan and green lines corresponding to the fits to 5, 6 and 7 peaks respectively. The legend shows the fitted constants of the linear fit.}
    \label{DECAYPLOTS}
   
\end{figure}

Figures \ref{WAVELETPLOTS}, \ref{FOURIERPLOTS}, and \ref{DECAYPLOTS} show the plots for the wavelet analysis, Fourier transform, and decay rates respectively for each simulation of the parameter study.

%%%%%%%%%%%%%%%%%%%%%%%%%%%%%%%%%%%%%%%%%%%%%%%%%%%%%%%%
%%%%%%%%%%%%%%%%%%%%%%%%%%%%%%%%%%%%%%%%%%%%%%%%%%%%%%%%

%%%%%%%%%%%%%%%%%%%%%%%%%%%%%%%%%%%%%%%%%%%%%%%%%%%%%%%%
%%%%%%%%%%%%%%%%%%%%%%%%%%%%%%%%%%%%%%%%%%%%%%%%%%%%%%%%

%%%%%%%%%%%%%%%%%%%%%%%%%%%%%%%%%%%%%%%%%%%%%%%%%%%%%%%%
%%%%%%%%%%%%%%%%%%%%%%%%%%%%%%%%%%%%%%%%%%%%%%%%%%%%%%%%

\bibliography{bib}{}
\bibliographystyle{aasjournal}

%%%%%%%%%%%%%%%%%%%%%%%%%%%%%%%%%%%%%%%%%%%%%%%%%%%%%%%%
\end{document}